\documentclass[journal,onecolumn]{IEEEtran}
\usepackage{cite}

\ifCLASSINFOpdf
\else
\fi
\usepackage{amsmath,amsthm,amssymb}
\allowdisplaybreaks

\usepackage{array}

\usepackage{stfloats}
\usepackage{url}

\usepackage[caption=false,font=footnotesize]{subfig}
\usepackage{enumitem}

\usepackage[pdftex]{graphicx}
\newtheorem{theorem}{Theorem}
\newtheorem{definition}{Definition}
\newtheorem{lemma}[theorem]{Lemma}

\newtheorem{corollary}[theorem]{Corollary}
\newtheorem{remark}{Remark}
\newtheorem{example}{Example}

\begin{document}
\title{On Vector Linear Solvability of non-Multicast Networks}
%
%
%

\author{Niladri Das and
        Brijesh Kumar Rai
\thanks{Niladri Das and Brijesh Kumar Rai are with the Department
of Electronics and Electrical Engineering, Indian Institute of Technology Guwahati,
Assam, 781039 India e-mail: d.niladri@iitg.ac.in and bkrai@iitg.ac.in.}
\thanks{Manuscript received on ----- -------}
\thanks{Part of this paper has been presented at the 2018 IEEE International Symposium on Information Theory and Its Applications (ISITA), Singapore \cite{isita}.}}

%
%

%

\maketitle

\begin{abstract}
Vector linear solvability of multicast networks neither depends upon the characteristic of the finite field nor on the dimension of the vector linear network code. However, vector linear solvability of non-multicast networks depends upon both the characteristic of the finite field and the dimension of the code. In the literature, the dependency on the characteristic of the finite field and the dependency on the dimension have been studied separately. 
In this paper, we show the interdependency between the characteristic of the finite field and the dimension of the vector linear network code that achieves a vector linear network coding (VLNC) solution. Towards this end, for any given network $\mathcal{N}$, we define $P(\mathcal{N},d)$ as the set of all characteristics of finite fields over which the network $\mathcal{N}$ has a $d$-dimensional VLNC solution. To the best of our knowledge, for any network $\mathcal{N}$ shown in the literature, if $P(\mathcal{N},1)$ is non-empty, then $P(\mathcal{N},1) = P(\mathcal{N},d)$ for any positive integer $d$. We show that, for any two non-empty sets of primes $P_1$ and $P_2$, there exists a network $\mathcal{N}$ such that $P(\mathcal{N},1) = P_1$, but $P(\mathcal{N},2) = \{P_1,P_2 \}$. We also show that there are networks exhibiting a similar advantage (the existence of a VLNC solution over a larger set of characteristics) if the dimension is increased from $2$ to $3$. However, such behaviour is not universal, as there exist networks which admit a VLNC solution over a smaller set of characteristics of finite fields when the dimension is increased.
%
%
Using the networks constructed in this paper, we further demonstrate that: (i) a network having an $m_1$-dimensional VLNC solution and an $m_2$-dimensional VLNC solution may not have a $m_1 + m_2$-dimensional VLNC solution; (ii) there exist a class of networks exhibiting some advantage in using non-commutative rings as the source alphabet: the least sized non-commutative ring over which each network in the class has a scalar linear network coding (SLNC) solution is significantly lesser in size than the least sized finite field over which it has an SLNC solution.

\end{abstract}


%
\IEEEpeerreviewmaketitle

%
%
%
%

 

\section{Introduction}
We consider the problem of communicating a set of vectors from the sources to the receivers over a wired network. Ahlswede \textit{et al.} introduced the concept of network coding where intermediate nodes of a network can compute and forward functions of incoming messages rather than sending intrinsic source messages \cite{ahlswede}. Network coding has been shown to perform better than routing in terms of achieving a higher data rate in various communication problems over a network. In particular, for a class of communication networks termed as multicast networks, the min-cut upper bound on the achievable data rates is achievable using network coding, but, in general, is not achievable using routing.

In this paper, we study the problem of communicating vectors over a network under the setting of a restricted version of network coding, called linear network coding (LNC). In LNC, all the nodes in the network can use only linear functions to encode or decode messages. Li \textit{et al.} showed that, over a sufficiently large finite field, the capacity of any multicast network is achievable using a form of LNC called scalar LNC (SLNC). In SLNC, the vector of symbols forwarded by a node (or decoded by a terminal) is a linear combination of the vector of symbols received by the node. Jaggi \textit{et al.} devised a polynomial time algorithm for designing capacity achieving scalar linear network codes in multicast networks \cite{jaggi}.  Ho \textit{et al.} showed that these capacity achieving codes could be deployed randomly with the success probability tending to $1$ as the size of the finite field is increased \cite{ho}.  

Scalar linear solvability of multicast networks has been shown to be dependent upon the size of the finite field. For any positive integer $n$, it has been shown that there exists a multicast network which has an SLNC solution if the size of the finite field is greater than or equal to $n$, but has no SLNC solution if the size of the finite field is less than $n$. In 2015, Sun \textit{et al.} showed that there exists an infinite number of networks where each network has an SLNC solution over some finite field but does not have an SLNC solution over a larger finite field \cite{sun1}. In references \cite{sun1} and \cite{sunbase}, the authors showed that not only the size but also the order and the associated coset numbers of the subgroups of the multiplicative group of the finite field affects the existence of an SLNC solution.

Vector LNC (VLNC) is a generalization of SLNC: instead of using scalar quantities as coefficients to linearly combine incoming vectors, VLNC uses square matrices as coefficients. In a $d$-dimensional VLNC over a finite field $\mathbb{F}_q$, each source forwards a $d$-length vector over a finite field $\mathbb{F}_q$ ($d$ symbols from $\mathbb{F}_q$), each edge carries $d$-length vectors over $\mathbb{F}_q$, each non-source node upon receiving the incoming vectors (one $d$-length vector from each incoming edge), multiplies each vector by a $d\times d$ matrix (called a local coding matrix), and sums the resultant vectors to form a new vector, which it can forward through outgoing edges, or store in its memory as a vector it has decoded. We refer the number $d$ in a $d$-dimensional VLNC as the message dimension.



Ebrahimi \textit{et al.} presented an efficient algorithm to design vector linear network codes (the set of local coding matrices) that achieves VLNC solutions for multicast networks \cite{ebrahimi}. It has been shown in \cite{j-linear} and \cite{sun2} that if a network has an SLNC solution over a finite field $\mathbb{F}_{q^L}$, it has an $L$-dimensional VLNC solution over $\mathbb{F}_q$ as well. The work of Ebrahimi \textit{et al.} in \cite{ebrahimi} indicated that there may exist a multicast network which has an $L$-dimensional VLNC solution over a finite field $\mathbb{F}_q$ but has no SLNC solution over any finite field whose size is less than or equal to $q^L$. This conjecture was settled by Sun \textit{et al.} in \cite{sun2} by showing explicit instances of networks exhibiting such behaviour. Subsequently, in \cite{etzion}, Etzion \textit{et al.} showed networks in which a much larger gap between the least integer $q^L$ such that the network has an $L$-dimensional VLNC solution over a finite field $\mathbb{F}_q$, and the size of the smallest finite field over which the network has an SLNC solution, is observed. In \cite{sun2}, Sun \textit{et al.}  showed that there exists a multicast network which has a $4$-dimensional VLNC solution  over $\mathbb{F}_2$, but has no $5$-dimensional VLNC solution over $\mathbb{F}_2$ (the network has a $5$-dimensional VLNC solution over $\mathbb{F}_{2^4}$, so the solution is not dependent upon the characteristic of the finite field). 

%

Non-multicast networks or multi-source multi-terminal networks act very differently to multicast networks. SLNC has been shown to be insufficient to achieve the capacity of non-multicast networks. M{\'{e}}dard \textit{et al.} showed that there exists a non-multicast network: the M-network, which has no SLNC solution over any finite field, but has a $2$-dimensional VLNC solution over all finite fields \cite{medard}. Dougherty \textit{et al.}, in \cite{matroid}, showed that the M-network has a $d$-dimensional VLNC solution if and only if $d$ is even. These two results were further generalized by Das \textit{et al.}, in \cite{das}, to show that for any positive integer $m \geq 2$, there exists a network which has a VLNC solution if the message dimension is a positive integer multiple of $m$, but does not have a VLNC solution otherwise. Das \textit{et al.} also showed in \cite{das2} that for any positive integer $m\geq 3$, there exists a network which has a VLNC solution if the message dimension is greater than or equal to $m-1$, but does not have a VLNC solution otherwise. However, VLNC has been also shown to be insufficient to achieve a solution of all non-multicast networks \cite{insuff}. Dougherty \textit{et al.} showed that more general forms of LNC defined over non-field rings and modules may also fail to achieve a solution of such networks \cite{insuff}. Reference \cite{non-linear} shows an infinite class of networks which admit a non-linear solution, but has no SLNC or VLNC solution over any ring or module.

One of the distinctions between multicast networks and non-multicast networks is that in the latter, the existence of an SLNC or VLNC solution depends upon the characteristic of the field.  It has been shown in \cite{poly} that for any set of polynomials with integer coefficients, there exists a network which has an SLNC solution over a finite field if and only if the set of polynomials have a common root over the field. They also showed that the set of all characteristics of finite fields over which a network has an SLNC solution is either finite or co-finite. Rai \textit{et al.} in \cite{rai}, by showing a connection between solvability of sum-networks and solvability of non-multicast networks, proved that for any finite/co-finite set of primes there exists a network which has a VLNC solution if and only if the characteristic of the finite field belongs to the given set of primes. For any finite/co-finite sets of primes, Connelly \textit{et al.} in \cite{non-linear} showed an instance of a non-multicast network that admits an SLNC solution if and only if the characteristic of the finite field belongs to the given set of primes.

Most works on LNC in the literature consider the source alphabet as a finite field. Recently LNC over rings and modules has been studied extensively in three companion papers: \cite{connelly1}, \cite{connelly2}, and \cite{LNCrings}. In \cite{connelly1}, the authors investigated whether SLNC over commutative rings has any advantage over SLNC over finite fields. Commutative rings are more general than finite fields. However, the authors showed that if a network has an SLNC solution over a finite commutative ring which is not a field, then the network also has an SLNC solution over a finite field whose size is less than or equal to the size of the commutative ring. This indicates that if the goal is to achieve an SLNC solution over the least sized alphabet, then there is no need to look for non-field commutative rings. Moreover, this result also shows that generality of the alphabet structure does not necessarily imply any advantage in terms of existence of a SLNC solution over a lesser sized alphabet. In \cite{connelly2}, it was shown that all networks having a VLNC solution over some finite field also have an SLNC solution over some finite ring; if a network has no SLNC solution over any finite field but has a VLNC solution over some finite field, then the network has an SLNC solution over some non-commutative ring. They showed an infinite class of networks which has an SLNC solution over some non-commutative ring, but has no SLNC solution over any commutative ring. Additionally, the size of such a non-commutative ring must be greater than or equal to $16$, and this bound is achieved in the M-network shown in \cite{medard}. In \cite{LNCrings}, the authors showed that the linear coding capacity of any network over finite fields is greater than or equal to its linear coding capacity over rings and modules. They also showed that linear coding capacity over a finite field depends only on the characteristic of the finite field. In the following sub-section, we describe the notations used in this paper.


%

\subsection{Notations}
For the rest of this paper, unless otherwise mentioned, a network would indicate a non-multicast network. For any given network $\mathcal{N}$, we define $P(\mathcal{N},d)$ as the set of all primes such that the network $\mathcal{N}$ has a $d$-dimensional VLNC solution over a finite field if and only if the characteristic of the finite field belongs to $P(\mathcal{N},d)$. A network $\mathcal{N}$ has no $d$-dimensional VLNC solution over any finite field if and only if $P(\mathcal{N},d) = \emptyset$.


The notation $q$ may denote any positive integer, however in the context of a finite field $\mathbb{F}_{q}$, $q$ denotes a power of a prime. We use the notation $\mathbb{P}$ to denote the set of all prime numbers. A vector over a finite field $\mathbb{F}_q$ indicates that the components of the vector belong to $\mathbb{F}_q$. $\mathbb{F}_q^d$ denotes the set of $d$-length vectors over $\mathbb{F}_q$. We use the notation $\mathbb{Z}^+$ to denote the set of all positive integers. Next, we describe the contributions of this paper.

\subsection{Our Contributions}
To the best of our knowledge, for all networks shown in the literature, if $P(\mathcal{N},1) \neq \emptyset$, then $P(\mathcal{N},1) = P(\mathcal{N},d)$ for any $d \in \mathbb{Z}^+$ (\textit{i.e.} if a network $\mathcal{N}$ has an SLNC solution if and only if the characteristic of the finite field belongs to a certain set of primes, then it also has a VLNC solution for any message dimension if and only if the characteristic of the finite field belongs to the same set of primes). In fact, to the best of our knowledge, for all networks shown in the literature, if $\mathcal{N}$ has both a $d_1$-dimensional VLNC solution and a $d_2$-dimensional VLNC solution, then $P(\mathcal{N},d_1) = P(\mathcal{N},d_2)$.

We first show that, for any two non-empty sets of primes $P_1$ and $P_2$, there exists a network $\mathcal{N}$ such that $P(\mathcal{N},1) = P_1$, but $P(\mathcal{N},2) = \{P_1,P_2 \}$ (Theorem~\ref{main:1a}). This shows that if a network has both an SLNC solution and a VLNC solution, then the VLNC solution may exist over a larger set of characteristics of finite fields. We then show that, for any two non-empty sets of primes $P_1$ and $P_2$, there exists a network $\mathcal{N}$ such that $P(\mathcal{N},2) = P_1$, but $P(\mathcal{N},3) = \{P_1,P_2 \}$ (Theorem~\ref{main:2}). This shows that there exist networks in which if the message dimension is increased from $2$ to $3$, the set of characteristics of finite fields over which the network has a VLNC solution gets larger.


The results mentioned in the above paragraph may indicate that a higher message dimension is superior to a lower message dimension in terms of achieving a VLNC solution over a larger set of characteristics of finite fields. However, we also show counter-examples where this is not true. We show that, for any two non-empty sets of primes $P_1$ and $P_2$, there exists a network $\mathcal{N}$ for which $P(\mathcal{N},2) = \{P_1,P_2\}$, but $P(\mathcal{N},3) = P_2$ (Theorem~\ref{main:4}). This shows that there exist networks in which if the message dimension is increased by $1$, the size of the set of characteristics of finite fields over which the network has a VLNC solution gets smaller.


Using the networks constructed in this paper, we further show two more results. First, we show that there exists a network which has a $2$-dimensional VLNC solution and a $3$-dimensional VLNC solution, but has no $5$-dimensional VLNC solution (Theorem~\ref{main:5}). This shows that a network having an $m_1$-dimensional VLNC solution and an $m_2$-dimensional VLNC solution may not have a $m_1 + m_2$-dimensional VLNC solution. 

Second, we show that, for any prime $p$, there exists a network which has an SLNC solution over a non-commutative ring of size $16$, but has no SLNC solution over any finite field whose size is less than $p$, and has a SLNC solution over a finite field of size $p$ (Theorem~\ref{main:7}). So for $p > 16$, these networks have an SLNC solution over a non-commutative ring whose size is strictly less than the size of any finite field over which also an SLNC solution exists. To the best of our knowledge, this is the first result that explicitly shows the superiority of non-commutative rings in terms of size.

It has been shown in \cite{connelly1} that the least sized commutative ring over which a network has an SLNC solution is a field. Hence, the above result is true for all commutative rings, \textit{i.e.}, for any $p>16$ there exists a network which admits an SLNC over a non-commutative ring whose size is less than any commutative ring over which the network admits an SLNC solution.

\subsection{Organization of the paper}
In Section~\ref{sec1}, we re-produce the standard definitions related to VLNC. In Section~\ref{sec2}, we introduce the network constructions that will be used to establish the contributions of this paper. In Section~\ref{sec3}, we present the main results of the paper. In Section~\ref{sec4}, we conclude the paper. The proofs of most of the theorems and lemmas of Section~\ref{sec2} are deferred to Appendix. Some of the proofs require the concept of polymatroid algebra, and hence, for the sake of completeness, in the appendix we also give an introduction to discrete polymatroids and its connection to linear solvability of networks.
\section{Vector Linear Network Coding}\label{sec1}
A network is represented by a directed acyclic graph $G(V,E)$ where $V$ is the set of nodes and $E \subseteq V \times V$ is the set of edges. Three subsets of $V$ are defined: the set of sources $S$, the set of terminals $T$, and the set containing rest of the nodes $V^\prime$. Without loss of generality (w.l.o.g.), it is assumed that the sets $S, V^\prime,$ and $T$ are disjoint and partition $V$. Each source generates a $d$-length vector (called a message) which is uniformly distributed over $\mathbb{F}_q^d$. Any vector generated by a source is independent of the other vectors generated by other sources. Each terminal wants to receive the vectors generated by a subset of the sources. W.l.o.g., it is assumed that the sources have no incoming edges, and the terminals have no outgoing edges. Each edge carries an element from $\mathbb{F}_q^d$. A vector carried by an edge  is either a linear function of the messages generated by the tail node of the edge (if the tail node is a source), or a linear function of the symbols carried by the edges incoming to the tail node of the edge. A vector computed by a terminal is a linear function of the vectors carried by the edges incoming to the terminal. 

To compute these linear functions, for each: adjacent edge pair, source-edge pair where the source is the tail node of the edge, edge-terminal pair where the terminal is the head node of the edge, a $d \times d$ matrix belonging to $\mathbb{F}_q^{d\times d}$ is assigned. Each of these matrices is called a local coding matrix, and their collection is called a vector linear network code of $d$ message dimension.

For any edge $e \in E$, if the tail node of $e$ is a source $s$, then the vector carried by $e$ is equal to $A_{\{s,e\}}x_s$ where $A_{\{s,e\}}$ is the local coding matrix for the source-edge pair $(s,e)$, and $x_s$ is the vector generated by $s$. For any $e \in E$, if the tail node of $e$ is $v \in V^\prime$, then the vector carried by $e$ is equal to $\sum_{e^\prime \in In(v)} A_{\{e^\prime,e\}}y_{e^\prime}$ where $In(v)$ is the set of all edges whose head node is $v$,  $A_{\{e^\prime,e\}}$ is the local coding matrix for the adjacent edge pair $(e^\prime,e)$, and $y_{e^\prime}$ is the vector carried by the edge $e^\prime$. If a terminal $t$ computes a vector $x_t$, then $x_t = \sum_{e \in In(t)} A_{\{e,t\}}x_e$ where $In(t)$ is the set of all edges whose head node is $t$,  $A_{\{e,t\}}$ is a local coding matrix for the edge-terminal pair $(e,t)$, and $y_{e}$ is the vector carried by the edge $e$.

If all terminals are successful in retrieving their demanded messages, then the network is said to have a $d$-dimensional vector linear network coding (VLNC) solution over $\mathbb{F}_q$. The positive integer $d$ is called the message dimension or the vector dimension or the dimension of the vector linear network code. A vector linear network code of $1$ message dimension is called a scalar linear network code, and an $1$-dimensional VLNC solution is called a scalar linear network coding (SLNC) solution. If a network has a VLNC solution for some message dimension over $\mathbb{F}_q$, then the network is said to have a VLNC solution over $\mathbb{F}_q$. 
\section{Constituent Networks}\label{sec2}
In this section, we present networks $\mathcal{N}_1$ and $\mathcal{N}_2$, which exhibit the property that the set of characteristics over which a VLNC solution exists varies with the message dimension. These two networks are in turn constructed using four other intermediate networks: the M-network (shown in \cite{medard}), a generalization of the M-network (shown in \cite{das}), Char-$m$ network (shown in \cite{non-linear} and \cite{connelly2}), and the Char-$q$-$s$ network (our contribution). 
\begin{figure*}[!t]
\begin{center}
\subfloat[The M-network reproduced from \cite{medard}. We denote this network by $\mathcal{M}_2$.]{\includegraphics[width=0.33\textwidth]{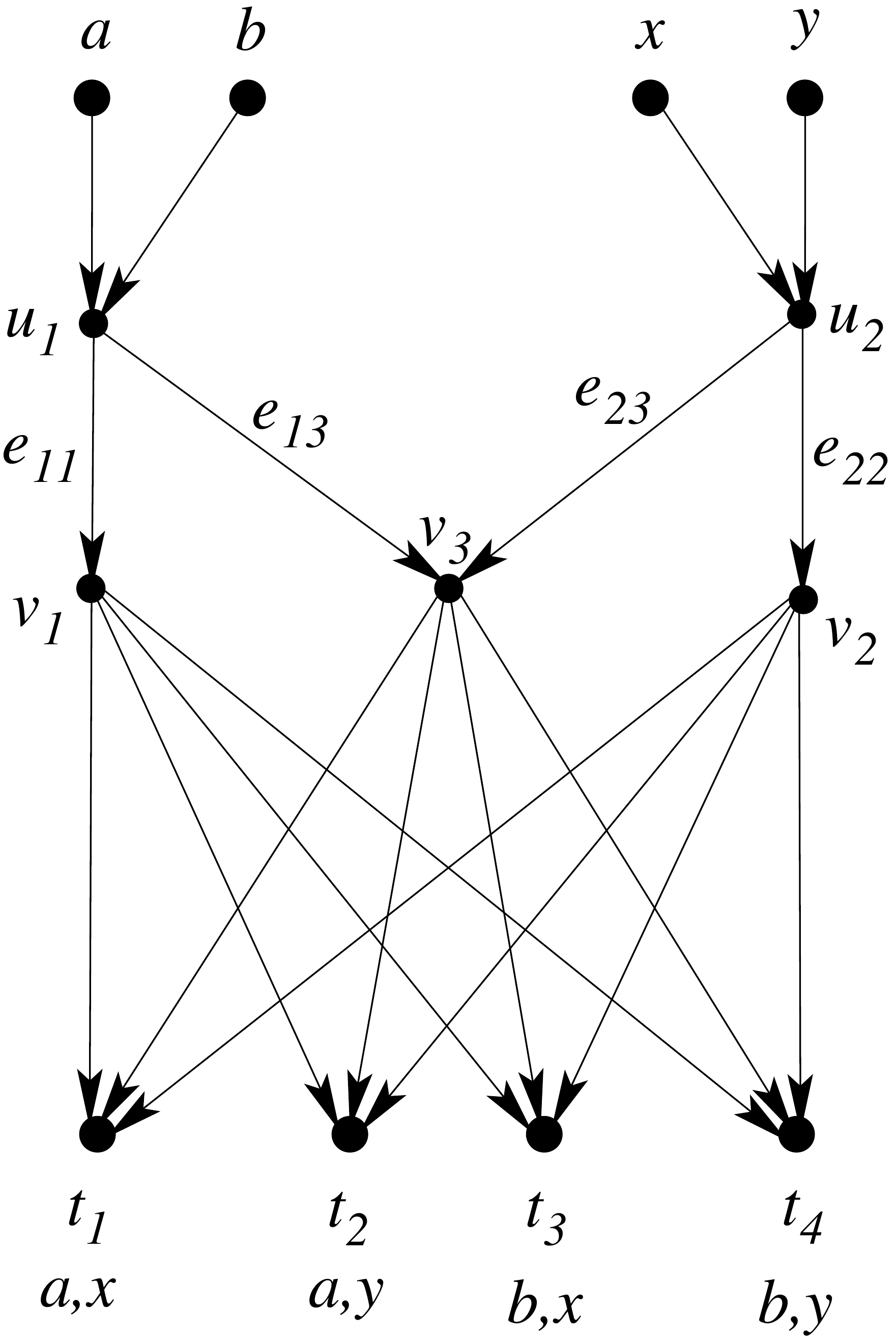}%
\label{fig:mnetwork}}
\hfil
\subfloat[The generalized M-network for $m=3$ reproduced from \cite{das}. We denote this network by $\mathcal{M}_3$.]{\includegraphics[width=0.44\textwidth]{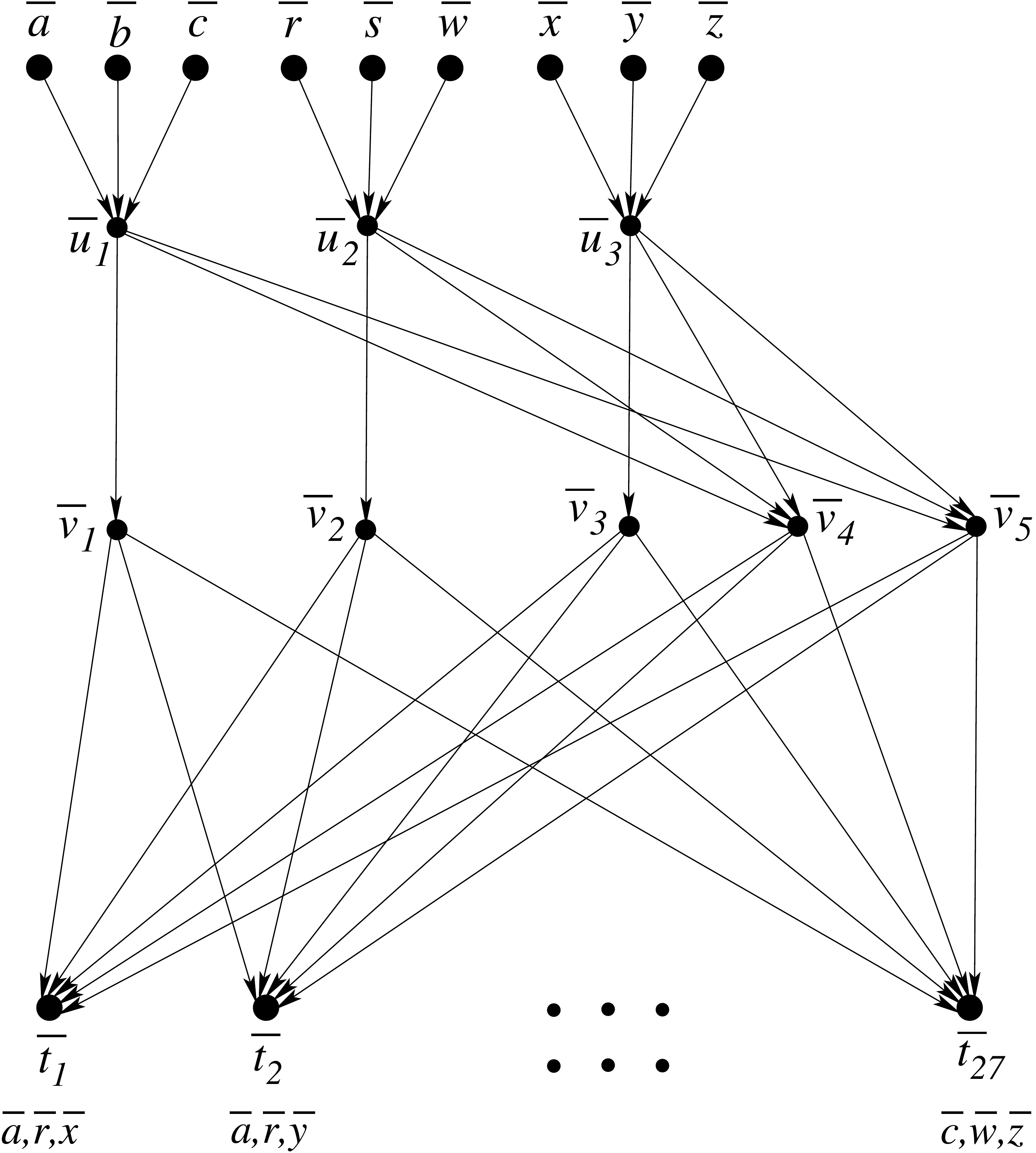}%
\label{fig:genmnetwork}}
\caption{The M-network and the generalized M-network for $m=3$.}
\label{fig_sim}
\end{center}
\end{figure*}
\subsection{M-network}\label{m_2}
The M-network was first shown in \cite{medard}. We denote the M-network by $\mathcal{M}_2$. We have reproduced this network in Fig.~\ref{fig:mnetwork}. The set of edges and vertices of the network are listed below (the labelling is different from \cite{medard}).
\begin{IEEEeqnarray*}{l}
S = \{a,b,x,y\}, V^\prime = \{u_1,u_2,v_1,v_2,v_3\}, T = \{t_i|1\leq i\leq 4\}.\\
E = \{(a,u_1), (b,u_1), (x,u_2), (y,u_2), (u_1,v_1), (u_1,v_3), (u_2,v_2), (u_2,v_3) \} \cup \{(v_i,t_j)|i=1,2,3, j=1,2,3,4\}.
\end{IEEEeqnarray*}
Let the message vector generated by a source be denoted by the same label as the source. Each terminal demands messages from a unique tuple of two sources, but none can demand $\{a,b\}$ or $\{x,y\}$; so there are $4$ possible demands, and hence $4$ terminals. The demands of the terminals are shown in Fig.~\ref{fig:mnetwork} below the terminal labels.

The following result has been proven in \cite{matroid}.
\begin{lemma}\label{Julylemma4}[Figure~1, \cite{medard}, and Theorem~V.10, \cite{matroid}]
For any $d \in \mathbb{Z}^+$, if $d$ is even then $P(\mathcal{M}_2,d) = \mathbb{P}$, else $P(\mathcal{M}_2,d) = \emptyset$.
\end{lemma}
%

%
%
\subsection{Generalized M-network for $m = 3$}\label{m_3}  
In reference \cite{das}, Das \textit{et al.} generalized the M-network. This generalization constructs one new network for each positive integer $m\geq 3$. We use the specific network that results for $m = 3$. Let us denote the generalized M-network for $m = 3$ by $\mathcal{M}_3$. This network is reproduced in Fig.~\ref{fig:genmnetwork}. The set of vertices and edges of this network is given below (the labelling is different from that of \cite{das}).
\begin{IEEEeqnarray*}{l}
S = \{\bar{a},\bar{b},\bar{c},\bar{r},\bar{s},\bar{w},\bar{x},\bar{y},\bar{z}\}, V^\prime =\{\bar{u}_i|i=1,2,3\}\cup \{\bar{v}_i|1\leq i\leq 5 \}, T = \{\bar{t}_i|1\leq i\leq 27\}.\\
E = \{(\bar{a},\bar{u}_1), (\bar{b},\bar{u}_1), (\bar{c},\bar{u}_1), (\bar{r},\bar{u}_2), (\bar{s},\bar{u}_2), (\bar{w},\bar{u}_2), (\bar{x},\bar{u}_3), (\bar{y},\bar{u}_3), (\bar{z},\bar{u}_3)\} \cup \{ (\bar{u}_i,\bar{v}_i), (\bar{u}_i,\bar{v}_4), (\bar{u}_i,\bar{v}_5) | i=1,2,3 \} \\ \hfill \cup  \{(\bar{v}_i,\bar{t}_j) | 1\leq i\leq 5, 1\leq j\leq 27\}\}.
\end{IEEEeqnarray*}
Let the message vector generated by a source be denoted by the same label as the source. Each terminal demands messages from a unique tuple of three sources: one source from $\{a,b,c\}$, one source from $\{r,s,w\}$, and one source from $\{x,y,z\}$; so there are $27$ possible tuples, and so there are $27$ terminals. 
The following result has been proven in \cite{das}.
\begin{lemma}\label{Julylemma3}[Theorem~1, \cite{das}]
For any $d \in \mathbb{Z}^+$, if $d$ is a multiple of $3$ then $P(\mathcal{M}_3,d) = \mathbb{P}$, else $P(\mathcal{M}_3,d) = \emptyset$.
\end{lemma}
%
%
\begin{figure*}[!t]
\centering
\subfloat[The Char-$m$ network reproduced from \cite{LNCrings}.]{\includegraphics[width=0.42\textwidth]{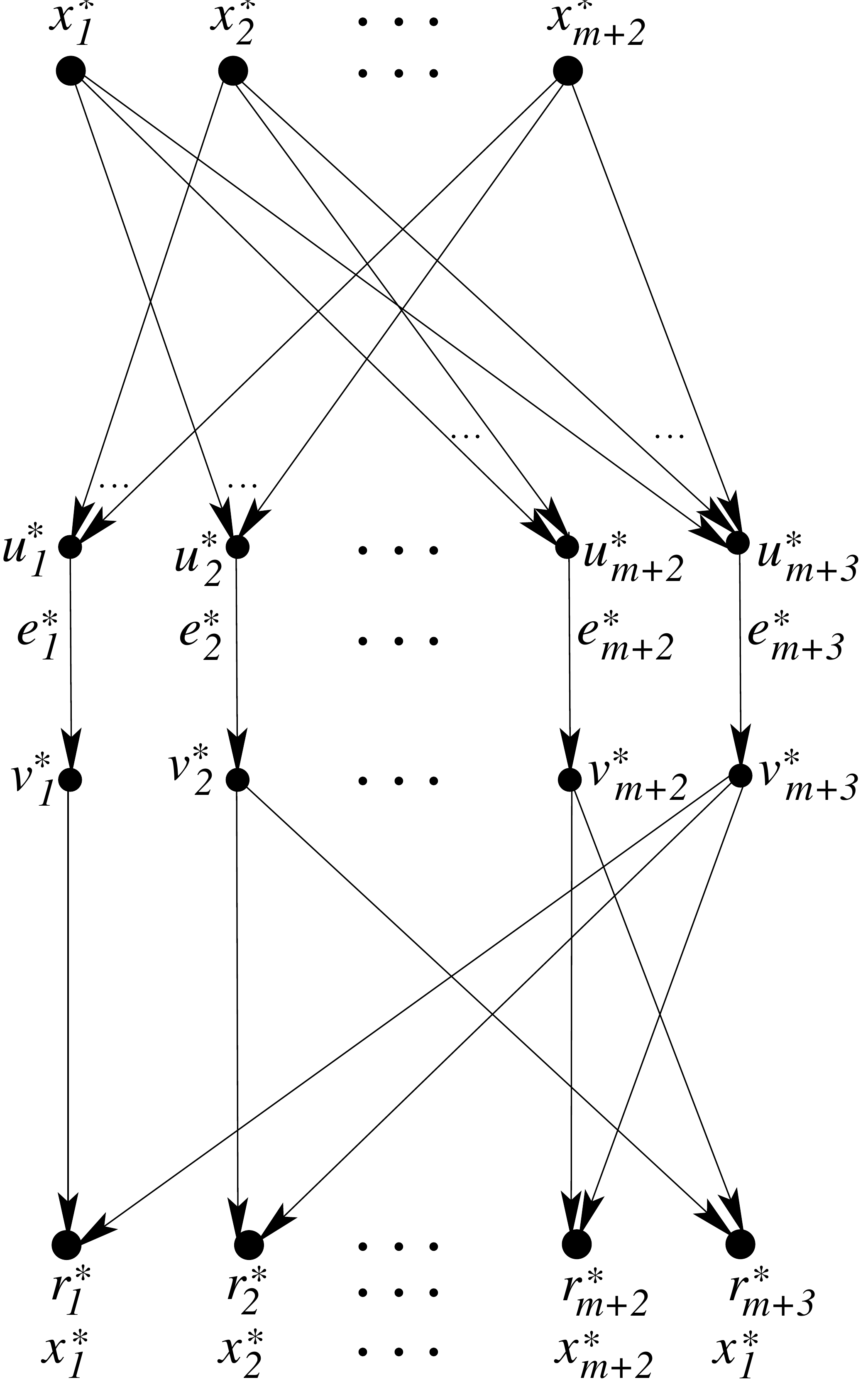}%
\label{fig:char}}
\hfil
\subfloat[The Char-$q$-$s$ network for $q = 2$. Note that the message generated by the source $s$ is not demanded by any terminal.]{\includegraphics[width=0.42\textwidth]{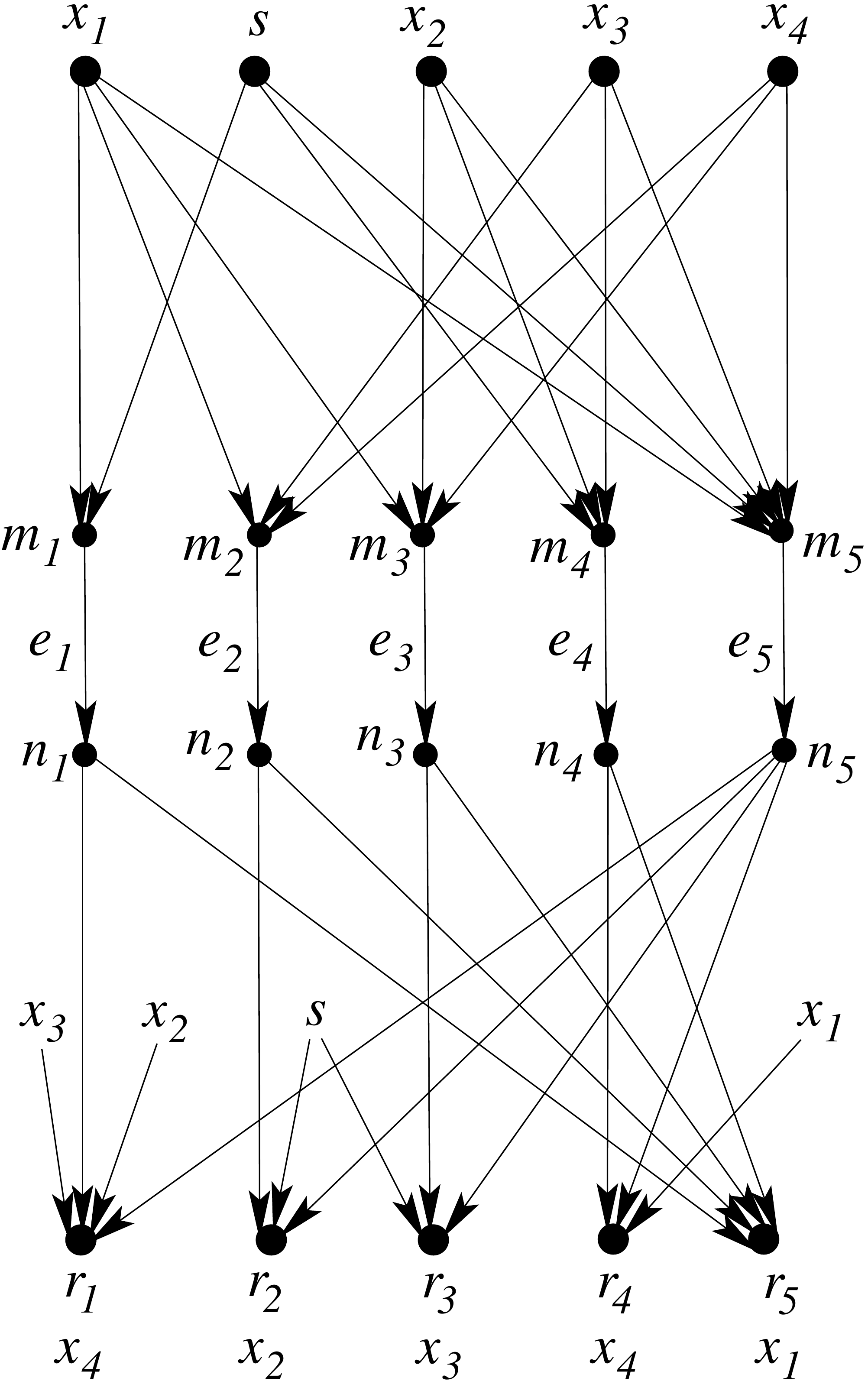}%
\label{char-qs}}
\caption{Networks Char-$m$ and Char-$2$-$s$. The demands of each terminal is shown below the terminal's label.}
\label{fig_sim2}
\end{figure*}
\subsection{Char-$m$ network}\label{char-q}
In \cite{LNCrings}, Connelly \textit{et al.}, for any integer $m \geq 2$, showed the Char-$m$ network. This network is a special case of a more general network construction shown in \cite{non-linear}. We have reproduced this network in Fig.~\ref{fig:char}. The set of edges and vertices of the network are listed below (the labelling is different from \cite{LNCrings}).
\begin{IEEEeqnarray*}{ll}
S =& \{x_1^*,x_2^*,\ldots,x_{m+2}^*\}, V^\prime = \{u_1^*,u_2^*,\ldots, u_{m+3}^*\} \cup \{v_1^*,v_2^*,\ldots, v_{m+3}^*\}, T = \{r_1^*,r_2^*,\ldots,r_{m+3}^*\}.\\
E =& \{(x_i^*,u_j^*) | 1\leq i\leq m+2, 1\leq j\leq m+3, i\neq j \}\cup \{e_i^* = (u_i^*,v_i^*)| 1\leq i\leq m+3 \} \cup\{(v_i^*,r_i^*) | 1\leq i\leq m+2 \}\\& \cup\{(v_{m+3}^*,r_i^*)| 1\leq i\leq m+2 \}\cup \{(v_i^*,r_{m+3}^*) | 2\leq i\leq m+2 \}.
\end{IEEEeqnarray*}
Let the message vector generated by a source be denoted by the same label as the source. The demands of the terminals are: $r_{i}^*$ for $1\leq i\leq m+2$ demands $x_i^*$, and $r_{m+3}^*$ demands $x_1^*$. 
The following result has been proven in \cite{non-linear}.
\begin{lemma}\label{Julylemma2}[Lemma~IV.7, \cite{non-linear}, and Lemma~III.1, \cite{LNCrings}]
For any $d \in \mathbb{Z}^+$, $P(\text{Char-}m,d) = \{p \in \mathbb{P} \;|\; p \text{ divides } m\}$.
\end{lemma}
\subsection{Char-$q$-$s$ network}\label{char}
%
%
%
%
Inspired by the Char-$m$ network, we construct the Char-$q$-$s$ network, where $q \geq 2$ is a positive integer and $s$ is a label of a source node. The source labelled by $s$ is distinguished from the rest because no terminal demands $s$. 
%
The Char-$q$-$s$ network for $q=2$ is shown in Fig.~\ref{char-qs}. 
The set of vertices and edges of the Char-$q$-$s$ network is as follows:
\begin{IEEEeqnarray*}{ll}
S =& \{s\} \cup \{x_1,x_2,\ldots,x_{q+2}\}, V^\prime = \{m_1,m_2,\ldots,m_{q+3}\} \cup \{n_1,n_2,\ldots,n_{q+3}\}, T = \{r_1,r_2,\ldots,r_{q+3}\}.\\
E =& \{(x_1,m_i)| 1\leq i\leq q+1\}\cup \{(s,m_i)| i=1, 4\leq i\leq q+3  \}\cup \{(x_i,m_j) | 2\leq i,j\leq q+2, i\neq j \}\\& \cup \{(x_i,m_{q+3}) | 1\leq i\leq q{+}2 \} \cup\{e_i = (m_i,n_i)| 1\leq i\leq q+3 \} \cup\{(n_i,r_i) | 1\leq i\leq q+2 \}\\& \cup\{(n_{q+3},r_i), (n_i,r_{q+3}) | 1\leq i\leq q+2 \}  \cup\{(x_i,r_1)| 2\leq i\leq q+1 \} \cup\{(x_1,r_{q+2})\} \cup \{(s,r_2), (s,r_3)\}.
\end{IEEEeqnarray*}
%
%
Let the message vector generated by a source be denoted by the same label as the source. The demands of the terminals are: $r_1$ demands $x_{q+2}$, $r_{i}$ for $2\leq i\leq q+2$ demands $x_i$, and $r_{q+3}$ demands $x_1$. Let the message vector generated by the source $x_i$ be also denoted by $x_i$, and the message vector generated by the source $s$ be also denoted by $s$.

The next three lemmas show that for any positive integer $d$, the set $P(\text{Char-}q\text{-}s,d)$ depends upon whether the edge $e_1$ carries a linear function of the vector generated by $s$. Let the vector carried by the edge $e_i$ be denoted by $y_{e_i}$ for $1\leq i\leq q+3$. Let us consider a $d$-dimensional vector linear network code over the Char-$q$-$s$ network. As per the description given in Section~\ref{sec1}, in terms of $d\times d$ matrices: $A_i$ for $i=1, 4\leq i\leq q+3$; $M_i$ for  $1\leq i\leq q+1$, $i=q+3$; $W_{(j,i)}$ for $2\leq j\leq q+3$, $2\leq i\leq q+2$, $j \neq i$; we have
\begin{IEEEeqnarray}{ll}
&y_{e_1} = M_1x_1 + A_1s\label{y1}\\
&y_{e_2} = M_2x_1 + \sum_{i=3}^{q+2} W_{(2,i)}x_i\label{y2}\\
&y_{e_3} = M_3x_1 + W_{(3,2)}x_2 + \sum_{i=4}^{q+2} W_{(3,i)}x_i\label{y3}\\
\text{for } 4\leq j\leq q+1:\;\, &y_{e_j} = M_jx_1 + A_js + \sum_{i=2,i\neq j}^{q+2} W_{(j,i)}x_i\label{yn}\\
&y_{e_{q+2}} = (A_{q+2})s + \sum_{i=2}^{q+1} W_{(q+2,i)}x_i\label{yq2}\\
&y_{e_{q+3}} = (M_{q+3})x_1 + (A_{q+3})s + \sum_{i=2}^{q+2} W_{(q+3,i)}x_i\label{yq3}
\end{IEEEeqnarray}
Now we show that if $A_1$ is not a zero matrix, then  $P(\text{Char-}q\text{-}s,d) = \{p \in \mathbb{P} | p \text{ divides } q \}$, and if $A_1$ is a zero matrix, then $P(\text{Char-}q\text{-}s,d) = \mathbb{P}$.
\begin{lemma}\label{Junelemma1}
For any $d \in \mathbb{Z}^+$, $P(\text{Char-}q\text{-}s,d) = \mathbb{P}$ if $A_1 = 0$.
\end{lemma}
\begin{IEEEproof}
We present an SLNC solution that holds over all finite fields when $A_1 = 0$. Consider a scalar linear network code where the middle edges carry the following vectors.
\begin{IEEEeqnarray*}{l}
y_{e_1} = x_1 \text{ and for } 2\leq j\leq q+3:\; y_{e_j} = \sum_{i=2,i\neq j}^{q+2} x_i
\end{IEEEeqnarray*}
We show this scalar linear network code forms a SLNC solution over all finite fields. $r_1$ can retrieve $x_{q+2}$ from $y_{e_{q+3}}$ as it receives messages $x_i$ for $2\leq i\leq q+1$ through direct edges. Terminal $r_i$ for $2\leq i\leq q+2$ retrieves $x_{i}$ by subtracting $y_{e_i}$ from $y_{e_{q+3}}$. And $r_{q+3}$ retrieves $x_1$ from $y_{e_1}$. 
\end{IEEEproof}
\begin{remark}
The the scalar linear network code shown in Lemma~\ref{Junelemma1} uses only addition operation: all encoding and decoding functions are either addition or subtraction (subtraction is addition with the additive inverse). So, when $A_1 = 0$, the Char-$q$-$s$ network has a SLNC solution over any group. We will use this property later in the paper.
\end{remark}
%
\begin{lemma}\label{Junelemma2}
Let $p$ be a prime such that $p$ does not divide $q$. Then, for any $d \in \mathbb{Z}^+$, $p \in P(\text{Char-}q\text{-}s,d)$ if and only if $A_1 = \mathbf{0}$.
\end{lemma}
\begin{IEEEproof}
The `if' part has been already proved in Lemma~\ref{Junelemma1}. We prove the `only if' part here. Say that over a finite field of characteristic $p$ ($p$ does not divide $q$), the Char-$q$-$s$ network has a $d$-dimensional VLNC solution for some positive integer $d$. Let $\mathbf{0}$ denote the $d\times d$ zero matrix, and $\mathbf{I}$ denote the $d \times d$ identity matrix. 
Due to the demands of the terminal $r_1$, from equations (\ref{y1}) and (\ref{yq3}), there exists $d\times d$ matrices $T_{11}$, $T_{12}$, and $T^\prime_{1j}$ for $2\leq j\leq q+1$, such that
\begin{equation}
(T_{11})y_{e_1} + (T_{12})y_{e_{q+3}} + \sum_{j=2}^{q+1} (T^\prime_{1j})x_{j} = x_{q+2}
\end{equation}
So we must have:
\begin{IEEEeqnarray}{l}
T_{11}M_1 + T_{12}M_{q+3} = \mathbf{0}\label{10f1}\\
T_{11}A_1 + T_{12}A_{q+3} = \mathbf{0}\label{10f2}\\
T_{12}W_{(q+3,q+2)} = \mathbf{I}\label{10f3}
\end{IEEEeqnarray}
Due to the demands of terminal $r_2$, from equations (\ref{y2}) and (\ref{yq3}),  there exists $d\times d$ matrices $T_{21}$, $T_{22}$, and $T_{2}^\prime$, such that
\begin{equation}
(T_{21})y_{e_2} + (T_{22})y_{e_{q+3}} + (T_{2}^\prime )s = x_{2}
\end{equation}
So we must have:
\begin{IEEEeqnarray}{ll}
&T_{21}M_2 + T_{22}M_{q+3} = \mathbf{0}\label{10f4}\\
&T_{22}W_{(q+3,2)} = \mathbf{I}\label{10f5}\\
\text{for } 3\leq i\leq q+2:\; &T_{21}W_{(2,i)} + T_{22}W_{(q+3,i)} = \mathbf{0}\label{10f6}
\end{IEEEeqnarray}
Due to the demands of terminal $r_3$, from equations (\ref{y3}) and (\ref{yq3}), there exists $d\times d$ matrices $T_{31}$, $T_{32}$, and $T_{3}^\prime$, such that
\begin{equation}
(T_{31})y_{e_3} + (T_{32})y_{e_{q+3}} + (T_{3}^\prime )s = x_{3}
\end{equation}
So we must have:
\begin{IEEEeqnarray}{ll}
&T_{31}M_3 + T_{32}M_{q+3} = \mathbf{0}\label{10f7}\\
&T_{31}W_{(3,2)} + T_{32}W_{(q+3,2)} = \mathbf{0}\label{10f8}\\
&T_{32}W_{(q+3,3)} = \mathbf{I}\label{10f9}\\
\text{for } 4\leq i\leq q+2:\; &T_{31}W_{(3,i)} + T_{32}W_{(q+3,i)} = \mathbf{0}\label{10f10}
\end{IEEEeqnarray}
Due to the demands of the terminal $r_j$ for $4\leq j\leq q+1$, from equations (\ref{yn}) and (\ref{yq3}), there exists $d\times d$ matrices $T_{j1}$ and $T_{j2}$ such that
\begin{equation}
(T_{j1})y_{e_j} + (T_{j2})y_{e_{q+3}} = x_{j}
\end{equation}
So we must have:
\begin{IEEEeqnarray}{ll}
&T_{j1}M_j + T_{j2}M_{q+3} = \mathbf{0}\label{10f11}\\
&T_{j1}A_j + T_{j2}A_{q+3} = \mathbf{0}\label{10f12}\\
&T_{j2}W_{(q+3,j)} = \mathbf{I}\label{10f13}\\
\text{for } 2\leq i \leq q+2, i\neq j:\; &T_{j1}W_{(j,i)} + T_{j2}W_{(q+3,i)} = \mathbf{0}\label{10f14}
\end{IEEEeqnarray}
Due to the demands of the terminal $r_{q+2}$, from equations (\ref{yq2}) and (\ref{yq3}), there exists $d\times d$ matrices $T_{(q+2)1}$, $T_{(q+2)2}$, and $T_{q+2}^\prime$, such that
\begin{equation}
(T_{(q+2)1})y_{e_{q+2}} + (T_{(q+2)2})y_{e_{q+3}} + (T_{q+2}^\prime )x_1 = x_{q+2}
\end{equation}
So we must have:
\begin{IEEEeqnarray}{ll}
&T_{(q+2)1}A_{q+2} + T_{(q+2)2}A_{q+3} = \mathbf{0}\label{10f15}\\
\text{for } 2\leq i\leq q+1:\; &T_{(q+2)1}W_{(q+2,i)} + T_{(q+2)2}W_{(q+3,i)} = \mathbf{0}\label{10f16}\\
&T_{(q+2)2}W_{(q+3,q+2)} = \mathbf{I}\label{10f17}
\end{IEEEeqnarray}
Due to the demands of the terminal $r_{q+3}$, from equations (\ref{y1})-(\ref{yq2}), there exists $d\times d$ matrices $Z_i$ for $1\leq i\leq q+2$ such that
\begin{equation}
(Z_1)y_{e_1} + (Z_2)y_{e_2} + \cdots + (Z_{q+2})y_{e_{q+2}} = x_1
\end{equation}
So we must have:
\begin{IEEEeqnarray}{ll}
&Z_1M_1 + Z_2M_2 + ... + Z_{q+1}M_{q+1} = \mathbf{I}\label{10f18}\\
&Z_1A_1 + Z_4A_4 + ... + Z_{q+2}A_{q+2} = \mathbf{0}\label{10f19}\\
\text{for } 2\leq i\leq q+2:\; 	&\sum_{j=2,j\neq i}^{q+2}  Z_jW_{(j,i)} = \mathbf{0}\label{10f20}
\end{IEEEeqnarray}
From equations (\ref{10f3}), (\ref{10f5}), (\ref{10f9}), (\ref{10f13}) and (\ref{10f17}), we get: $T_{i2}$ is invertible for $1\leq i\leq q+2$, and $W_{(q+3,i)}$ is invertible for $2\leq i\leq q+2$. Then, from equations (\ref{10f6}), (\ref{10f8}), (\ref{10f10}),  (\ref{10f14}) and (\ref{10f16}): $T_{i1}$ is invertible for $2\leq i\leq q+2$, and $W_{(j,i)}$ is invertible for $2\leq j,i\leq q+2$, $j \neq i$.

From equations (\ref{10f4}), (\ref{10f7}) and (\ref{10f11}), we have:
\begin{equation}
\text{for } 2\leq i\leq q+1:\; M_i = -T_{i1}^{-1}T_{i2}M_{q+3} \label{10f21}
\end{equation}
Substituting equation (\ref{10f21}) in equation (\ref{10f18}), we get:
\begin{equation}
Z_1M_1 - (Z_2T_{21}^{-1}T_{22} + \cdots + Z_{q+1}T_{(q+1)1}^{-1}T_{(q+1)2})M_{q+3} = \mathbf{I}\label{10f32}
\end{equation}
From equations (\ref{10f12}), and (\ref{10f15}), we have:
\begin{equation}
\text{for } 4\leq i\leq q+2: \quad A_i = -T_{i1}^{-1}T_{i2}A_{q+3} \label{10f22}
\end{equation}
Substituting equation (\ref{10f22}) in equation (\ref{10f19}), we get:
\begin{equation}
Z_1A_1 - (Z_4T_{41}^{-1}T_{42} + \cdots + Z_{q+2}T_{(q+2)1}^{-1}T_{(q+2)2})A_{q+3} = \mathbf{0}\label{10f33}
\end{equation}
From equations (\ref{10f6}), (\ref{10f8}), (\ref{10f10}), (\ref{10f14}) and (\ref{10f16}), we have:
\begin{equation}
\text{for } 2\leq j,i\leq q+2, j\neq i:\; W_{(j,i)} = -T_{j1}^{-1}T_{j2}W_{(q+3,i)}\label{10f23}
\end{equation}
Substituting equation (\ref{10f23}) in equation (\ref{10f20}), we have:
\begin{equation}
\text{for } 2\leq i\leq q+2:\; \sum_{j=2,j\neq i}^{q+2} Z_jT_{j1}^{-1}T_{j2}W_{(q+3,i)} = \mathbf{0}\label{10f24}
\end{equation}
Since $W_{(q+3,i)}$ for $2\leq i\leq q+2$ has been already shown to be invertible, we must have:
\begin{equation}
\text{for } 2\leq i\leq q+2:\; \sum_{j=2,j\neq i}^{q+2} Z_jT_{j1}^{-1}T_{j2} = \mathbf{0}\label{10f28}
\end{equation}
Expanding equation (\ref{10f28}) for each value of $2\leq i\leq q+2$, we have:
\begin{IEEEeqnarray}{l}
Z_3T_{31}^{-1}T_{32} + Z_4T_{41}^{-1}T_{42} + \cdots + Z_{q+2}T_{(q+2)1}^{-1}T_{(q+2)2} = \mathbf{0}\label{10f25}\\
Z_2T_{21}^{-1}T_{22} + Z_4T_{41}^{-1}T_{42} + \cdots + Z_{q+2}T_{(q+2)1}^{-1}T_{(q+2)2} = \mathbf{0}\label{10f26}\\
\qquad\qquad\qquad\vdots \qquad\qquad\qquad\qquad \vdots\\
Z_2T_{21}^{-1}T_{22} + Z_3T_{31}^{-1}T_{32} + Z_4T_{41}^{-1}T_{42} + \cdots + Z_{q+1}T_{(q+1)1}^{-1}T_{(q+1)2} = \mathbf{0}\label{10f27}
\end{IEEEeqnarray}
Substituting equation (\ref{10f27}) in equation (\ref{10f32}), we get
\begin{equation}
Z_1M_1 = \mathbf{I}\label{10f34}
\end{equation}
Adding the $q+1$ equations shown in equations (\ref{10f25})-(\ref{10f27}), i.e. by the operation $\sum_{i=2}^{q+2} \sum_{j=2,j\neq i}^{q+2} Z_jT_{j1}^{-1}T_{j2}$, we have:
\begin{equation}
q(Z_2T_{21}^{-1}T_{22} + Z_3T_{31}^{-1}T_{32} + Z_4T_{41}^{-1}T_{42} + \cdots + Z_{q+2}T_{(q+2)1}^{-1}T_{(q+2)2}) = \mathbf{0}\label{10f29}
\end{equation}
Since the characteristic of the finite field does not divide $q$, we must have $q \neq 0$ in the finite field. Then, from equation (\ref{10f29}), we must have:
\begin{equation}
Z_2T_{21}^{-1}T_{22} + Z_3T_{31}^{-1}T_{32} + Z_4T_{41}^{-1}T_{42} + \cdots + Z_{q+2}T_{(q+2)1}^{-1}T_{(q+2)2} = \mathbf{0}\label{10f30}
\end{equation}
For each value of $2\leq i\leq q+2$, subtracting equation (\ref{10f28}) from (\ref{10f30}), we get:
\begin{equation}
\text{for } 2\leq j\leq q+2:\; Z_jT_{j1}^{-1}T_{j2} = \mathbf{0}\label{10f31}
\end{equation}
Substituting the values set by equation (\ref{10f31}) in equation (\ref{10f33}), we get:
\begin{equation}
Z_1A_1 = \mathbf{0}\label{10f35}
\end{equation}
Since $Z_1$ is invertible due to equation (\ref{10f34}), we must have $A_1 = \mathbf{0}$. This proves the `only if' part.
\end{IEEEproof}
\begin{lemma}\label{Junelemma3}
If $A_i \neq \mathbf{0}$, then, for any $d \in \mathbb{Z}^+$, $P(\text{Char-}q\text{-}s,d) = \{p \in \mathbb{P} | p \text{ divides } q \}$.
\end{lemma}
\begin{IEEEproof}
Since $A_i \neq \mathbf{0}$, Lemma~\ref{Junelemma2} shows that if $p \in \mathbb{P}$ and $p$ does not divide $q$, then $p \notin P(\text{Char-}q\text{-}s,d)$ for any $d \in \mathbb{Z}^+$. So it must be that $P(\text{Char-}q\text{-}s,d) \subseteq \{p \in \mathbb{P} | p \text{ divides } q \}$. We now show that if $p \in \mathbb{P}$ and $p$ divides $q$, then $p \in P(\text{Char-}q\text{-}s,d)$ for any $d\in \mathbb{Z}^+$, which will show that $\{p \in \mathbb{P} | p \text{ divides } q \} \subseteq P(\text{Char-}q\text{-}s,d)$, thereby proving the lemma. Consider a scalar linear network code where the middle edges carry the following vectors.
\begin{IEEEeqnarray}{l}
y_{e_1} = x_1 + s\label{23f1}\\
y_{e_2} = x_1 + x_3 + \cdots + x_{q+2}\label{23f2}\\
y_{e_3} = x_1 + x_2 + x_4 + \cdots + x_{q+2}\label{23f3}\\
\text{for } 4\leq j\leq q+1:\; y_{e_j} = s + \sum_{i=1,i\neq j}^{q+2} x_i\label{23f4}\\
y_{e_{q+2}} = s + \sum_{i=2}^{q+1} x_i\label{23f5}\\
y_{e_{q+3}} = s + \sum_{i=1}^{q+2} x_i\label{23f6}
\end{IEEEeqnarray}
Terminal $r_1$ receives messages $x_i$ for $2\leq i\leq q+1$ through direct edges, $x_1 + s$ from $e_1$, and hence it can retrieve $x_{q+2}$ from $y_{e_{q+3}}$. Terminal $r_2$ receives $s$ from a direct edge, and hence it can subtract  $y_{e_2} + s$ from $y_{e_{q+3}}$ to receive $x_2$. Similarly, $r_3$ receives $x_3$. Terminal $r_i$ for $4\leq i\leq q+1$ receives $x_{i}$ by subtracting $y_{e_i}$ from $y_{e_{q+3}}$. Terminal $r_{q+2}$ receives $x_1$ from a direct edge, and hence it computes $x_{q+2}$ by subtracting $y_{e_{q+2}} + x_1$ from $y_{e_{q+3}}$. And since $q=0$ over the finite field, $r_{q+3}$ receives $x_1$ by the operation: $\sum_{i=1}^{q+2} y_{e_i}$, as
\begin{equation}
\sum_{i=1}^{q+2} y_{e_i} = (q+1)x_1 + qs + \sum_{j=2}^{q+2} qx_j = x_1\label{4ap1}
\end{equation}
\end{IEEEproof}
\subsection{Network $\mathcal{N}_1$}\label{n1}
\begin{figure*}
\centering
\includegraphics[width=0.6\textwidth]{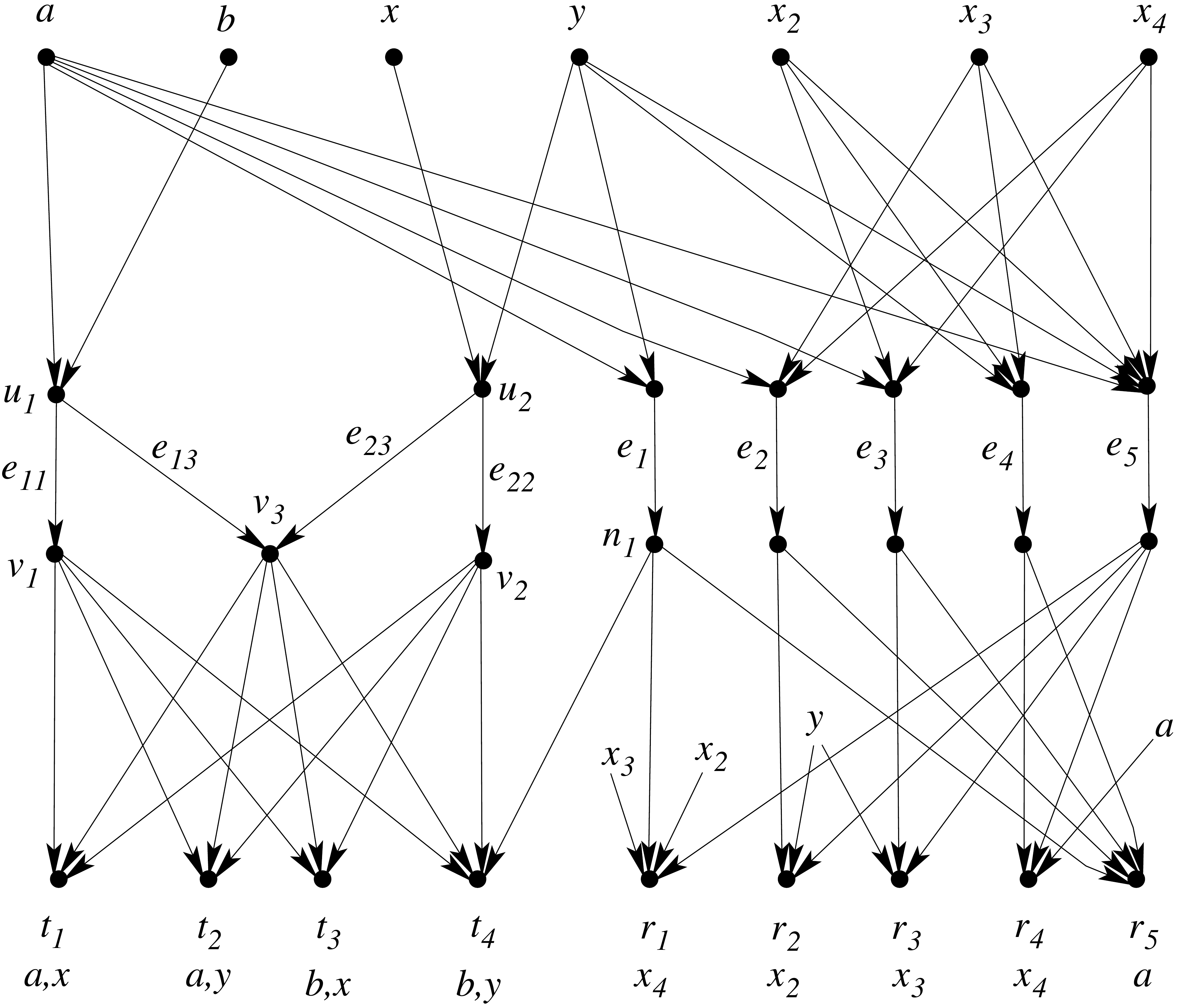}
\caption{The network $\mathcal{N}_1$ for $q = 2$. This network is a combination of the M-network, the Char-$2$-$y$ network, and an edge ($(n_1,t_4)$). The sources $a$ and $y$ are common to both of the M-network and the Char-$2$-$y$ network. The demands of the terminals are written below the label of the terminals.
}
\label{g1}
\end{figure*}
The network $\mathcal{N}_1$  is constructed by combining the M-network, the Char-$q$-$y$ network, and a new edge. $\mathcal{N}_1$ for the particular case of $q=2$ is shown in Fig.~\ref{g1}. 
Let the set of sources, intermediate nodes, terminals, and edges of the M-network (shown in Section~\ref{m_2}) be denoted by $S_{\mathcal{M}_2}$, $V^\prime_{\mathcal{M}_2}$,  $T_{\mathcal{M}_2}$, and $E_{\mathcal{M}_2}$ respectively; and let the set of sources, intermediate nodes, terminals,  and edges of the Char-$q$-$s$ network (shown in Section~\ref{char}) be denoted by $S_{\text{Char-$q$-$s$}}$, $V^\prime_{\text{Char-$q$-$s$}}$, $T_{\text{Char-$q$-$s$}}$, and $E_{\text{Char-$q$-$s$}}$ respectively. The set of vertices and edges of $\mathcal{N}_1$ is given below:
\begin{IEEEeqnarray*}{ll}
S = &S_{\mathcal{M}_2} \cup \{S_{\text{Char-$q$-$s$}}\setminus \{x_1,s\}\},\; V^\prime = V^\prime_{\mathcal{M}_2} \cup V^\prime_{\text{Char-$q$-$s$}},\; 
T = T_{\mathcal{M}_2} \cup T_{\text{Char-$q$-$s$}},\\
E = &E_{\mathcal{M}_2} \cup \{E_{\text{Char-$q$-$s$}}\setminus \{\{(x_1,m_i)| 1\leq i\leq q+1, i=q+3\}, (x_1,r_{q+2}), \{(s,m_i)| i=1, 4\leq i\leq q+3 \}, (s,r_2), (s,r_3) \} \}\\
&\cup \{\{(a,m_i)| 1\leq i\leq q+1, i=q+3\}, (a,r_{q+2}), \{(y,m_i)| i=1, 4\leq i\leq q+3 \}, (y,r_2), (y,r_3) \}  \cup \{(head(e_1),t_4))\}.
\end{IEEEeqnarray*}
It can be seen that the graph of $\mathcal{M}_2$ is a sub-graph of the graph of $\mathcal{N}_1$; the demands of the terminals of that belong to the sub-graph of $\mathcal{M}_2$ remain unchanged. Terminal $r_1$ demands $x_{q+2}$, terminal $r_{i}$ for $2\leq i\leq q+2$ demands $x_i$, and terminal $r_{q+3}$ demands $a$.  
The reason these two networks are connected as such is the following. It is known that the M-network does not have an SLNC solution; but we figured that if the terminal $t_4$ receives an extra symbol which is a linear function of $a$ and $y$, then it does have an SLNC solution. In $\mathcal{N}_1$, the terminal $t_4$ can have this extra symbol if the vector carried by $e_1$ is a linear combination of both $a$ and $y$. But, from  Lemma~\ref{Junelemma3} and we know that if  such is the case, then for the network to have an SLNC solution the characteristic of the finite field has to divide $q$, thus limiting the set of characteristics over which an SLNC solution exists.
%
%
\begin{lemma}\label{4Mlemma1}
For any odd positive integer $d$, $P(\mathcal{N}_1,d) = \{p \in \mathbb{P} | p \text{ divides } q\}$.
\end{lemma} 
The proof of this lemma is deferred to Appendix~\ref{app1}.
\begin{lemma}\label{Julylemma1}
For any even positive integer $d$, $P(\mathcal{N}_1,d) = \mathbb{P}$.
\end{lemma}
\begin{IEEEproof}
The following vector linear network code achieves a VLNC solution of the network for $d = 2$. Let the source vectors be: $a = [a_1 \;a_2], b = [b_1\;b_2], x = [\hat{x}_1\;\hat{x}_2 ], y = [y_1\;y_2]$, and for $2\leq i\leq q+2$: $x_i = [x_{i1}\; x_{i2}]$. Now, select local coding matrices such that: $e_{11} = [a_1\;b_1]$, $e_{13} = [a_2\;b_2]$, $e_{22} = [\hat{x}_1\;y_1]$, $e_{23} = [\hat{x}_2\;y_2]$, $e_1 = a$, for $2\leq i\leq q+3$: $e_i = \sum_{j=2, j \neq i}^{q+2} x_i$. It can be easily seen that using this vector linear network code all terminals can retrieve all of their demands. The VLNC solution for any other even dimension can be obtained by repeating this code.
\end{IEEEproof}
\textit{Intuition}: The M-network (sub-network of $\mathcal{N}_1$) already has a $2$-dimensional VLNC solution over all finite fields \cite{medard}. Hence, the terminal $t_4$ of the M-network part does not need any information from $e_1$ for it to compute its demands. The Char-$q$-$y$ sub-network of $\mathcal{N}_1$ has an SLNC solution over all finite fields when $e_1$ carries a linear function of only $a$ (Lemma~\ref{Junelemma1}). Since both of these two sub-networks have a $2$-dimensional VLNC solution over all finite fields, $\mathcal{N}_1$ has a $2$-dimensional VLNC solution over all finite fields.
\subsection{Network $\mathcal{N}_2$}\label{n2}
\begin{figure*}
\centering
\includegraphics[height=0.435\textheight]{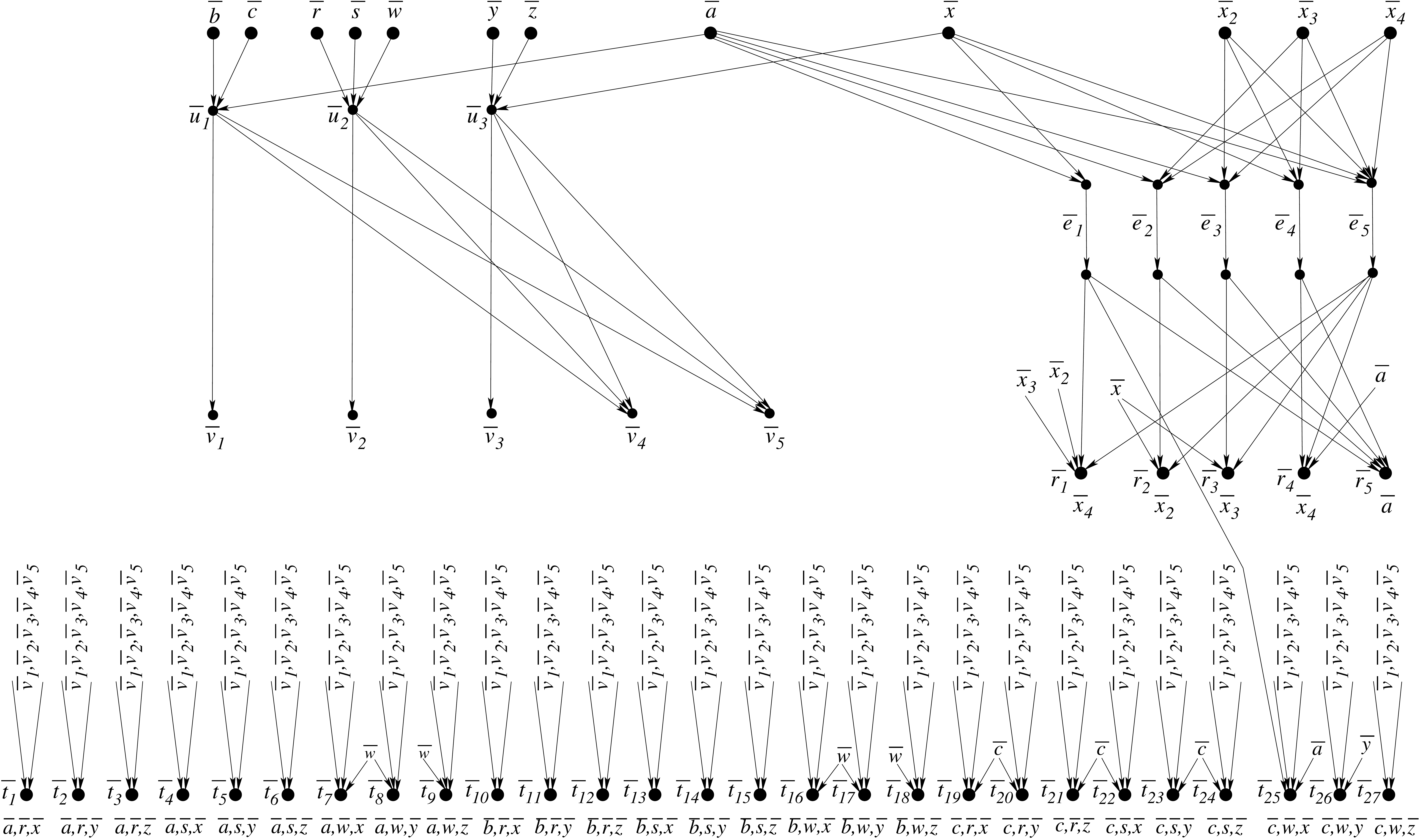}
\caption{The network $\mathcal{N}_2$ for $q^\prime = 2$. For some of the terminals, there is a direct edge connecting a source to the terminal, which we show by a truncated edge to maintain tidiness. For example, the terminal $\bar{t}_7$ has a direct edge $(\bar{w},t_7)$ connecting the source $\bar{w}$ and $\bar{t}_7$, but the complete edge has not been shown for the sake of clarity. Each intermediate nodes $\{\bar{v}_j|1\leq j\leq 5\}$ is connected to the terminal $\bar{t}_i$ for $1\leq i\leq 27$ by an edge $(\bar{v}_i,\bar{t}_j)$.
}
\label{g2}
\end{figure*}
The network $\mathcal{N}_2$ is constructed by joining together the generalized M-network for $m=3$ (reproduced in Section~\ref{m_3} as network $\mathcal{M}_3$), the Char-$q^\prime$-$\bar{x}$ network, and some additional edges. $\mathcal{N}_2$ for the particular case when $q^\prime=2$ is shown in Fig.~\ref{g2}. The set of vertices and edges of $\mathcal{N}_2$ is given below:
\begin{IEEEeqnarray*}{ll}
S &= \{\bar{a},\bar{b},\bar{c},\bar{r},\bar{s},\bar{w},\bar{x},\bar{y},\bar{z}\}\cup \{\bar{x}_2,\ldots,\bar{x}_{q^\prime +2}\},\\
V^\prime &=\{\bar{u}_i|i=1,2,3\}\cup \{\bar{v}_i|1\leq i\leq 5 \}\cup \{\bar{m}_1,\bar{m}_2,\ldots,\bar{m}_{q^\prime +3}\} \cup \{\bar{n}_1,\bar{n}_2,\ldots,\bar{n}_{q^\prime +3}\},\\
T &= \{\bar{t}_i|1\leq i\leq 27\}\cup  \{\bar{r}_1,\bar{r}_2,\ldots,\bar{r}_{q^\prime +3}\},\\
\IEEEeqnarraymulticol{2}{l}{\text{The set $E$ is written as an union of three sets $E_1, E_2, E_3$, where }E_1\text{ is a subset of the set of edges of $\mathcal{M}_3$}, E_2\text{ is the set of}}\\\IEEEeqnarraymulticol{2}{l}{\text{additional new edges}, \text{and $E_3$ is a subset of the set of edges of Char-$q^\prime$-$\bar{x}$, where the source $x_1$ of Char-$q^\prime$-$\bar{x}$ is replaced by}}\\\IEEEeqnarraymulticol{2}{l}{\text{the source $\bar{a}$ of $\mathcal{M}_3$, and the source $x_i$ for $2\leq i\leq q+2$ is replaced by $\bar{x}_i$.}}\\
E_1 &= \{(\bar{a},\bar{u}_1), (\bar{b},\bar{u}_1), (\bar{c},\bar{u}_1), (\bar{r},\bar{u}_2), (\bar{s},\bar{u}_2), (\bar{w},\bar{u}_2), (\bar{x},\bar{u}_3), (\bar{y},\bar{u}_3), (\bar{z},\bar{u}_3)\} \cup \{ (\bar{u}_i,\bar{v}_i), (\bar{u}_i,\bar{v}_4), (\bar{u}_i,\bar{v}_5) | i=1,2,3 \}\\
&\quad\cup  \{(\bar{v}_i,\bar{t}_j) | 1\leq i\leq 5, 1\leq j\leq 27\}\}\\
E_2 &= \{(\bar{w},\bar{t}_7), (\bar{w},\bar{t}_8), (\bar{w},\bar{t}_9), (\bar{w},\bar{t}_{16}), (\bar{w},\bar{t}_{17}), (\bar{w},\bar{t}_{18})\}
\cup \{(\bar{c},\bar{t}_{19}), (\bar{c},\bar{t}_{20}), (\bar{c},\bar{t}_{21}), (\bar{c},\bar{t}_{22}), (\bar{c},\bar{t}_{23}), (\bar{c},\bar{t}_{24})\} \cup \{(\bar{a},\bar{t}_{25})\}\\&\quad\cup \{(\bar{y},\bar{t}_{26})\} \cup \{(head(\bar{e}_1),\bar{t}_{25})\}\\
E_3 &= \{(\bar{a},\bar{m}_i)| 1\leq i\leq q^\prime +1, i=q^\prime +3\}\cup \{(\bar{x},\bar{m}_i)| i=1, 4\leq i\leq q^\prime +3  \}\cup \{(\bar{x}_i,\bar{m}_j) | 2\leq i,j\leq q^\prime +2, i\neq j \}\\
&\quad\cup \{(\bar{x}_i,\bar{m}_{q^\prime +3}) | 2\leq i\leq q^\prime +2 \} \cup\{\bar{e}_i = (\bar{m}_i,\bar{n}_i)| 1\leq i\leq q^\prime +3 \} \cup\{(\bar{n}_i,\bar{r}_i) | 1\leq i\leq q^\prime +2 \}\\
&\quad \cup\{(\bar{n}_{q^\prime +3},\bar{r}_i), (\bar{n}_i,\bar{r}_{q^\prime +3}) | 1\leq i\leq q^\prime +2 \}  \cup\{(\bar{x}_i,\bar{r}_1)| 2\leq i\leq q^\prime +1 \} \cup\{(\bar{a},\bar{r}_{q^\prime +2})\} \cup \{(\bar{x},r_2), (\bar{x},r_3)\}\\
%
%
E &= E_1 \cup E_2 \cup E_3.
\end{IEEEeqnarray*}
It can be seen that the graph of $\mathcal{M}_3$ is a sub-graph of the graph of $\mathcal{N}_2$; the demands of the terminals of that belong to this  sub-graph remain unchanged. Terminal $\bar{r}_1$ demands $\bar{x}_{q^\prime +2}$, terminal $\bar{r}_{i}$ for $2\leq i\leq q^\prime +2$ demands $\bar{x}_i$, and terminal $\bar{r}_{q^\prime +3}$ demands $\bar{a}$. 
%
%
%
We prove the following properties of $\mathcal{N}_2$. The proofs of all of these lemmas are given in Appendix~\ref{app2}.
\begin{lemma}\label{4Mlemma2}
$P(\mathcal{N}_2,1) = \emptyset$.
\end{lemma}
The proof is deferred to Appendix~\ref{app2.1}.

\begin{lemma}\label{4Mlemma3}
$P(\mathcal{N}_2,2) = \{p\in \mathbb{P}| p \text{ divides } q^\prime \}$.
\end{lemma}
The proof is deferred to Appendix~\ref{app2.2}.

\begin{lemma}\label{Julylemma6}
$P(\mathcal{N}_2,3) = \mathbb{P}$.
\end{lemma}
\begin{IEEEproof}
The following vector linear network code achieves a VLNC solution of the network for $d = 2$. Let the source vectors be: $\bar{a} = [\bar{a}_1 \;\bar{a}_2\;\bar{a}_3]$, $\bar{b} = [\bar{b}_1\;\bar{b}_2\;\bar{b}_3]$, $\bar{c} = [\bar{c}_1\;\bar{c}_2\;\bar{c}_3]$, $\bar{r} = [\bar{r}_1\;\bar{r}_2\;\bar{r}_3]$, $\bar{s} = [\bar{s}_1\;\bar{s}_2\;\bar{s}_3]$, $\bar{w} = [\bar{w}_1\;\bar{w}_2\;\bar{w}_3]$, $\bar{x} = [\hat{x}_1\;\hat{x}_2\;\hat{x}_3]$, $\bar{y} = [\bar{y}_1\;\bar{y}_2\;\bar{y}_3]$, $\bar{z} = [\bar{z}_1\;\bar{z}_2\;\bar{z}_3]$, and for $2\leq i\leq q^\prime +2$: $\bar{x}_i = [\bar{x}_{i1}\; \bar{x}_{i2}\; \bar{x}_{i3}]$. Now, select local coding matrices such that: $(\bar{u}_1,\bar{v}_1) = [\bar{a}_1 \;\bar{b}_1\;\bar{c}_1]$, $(\bar{u}_1,\bar{v}_4) = [\bar{a}_2 \;\bar{b}_2\;\bar{c}_2]$, $(\bar{u}_1,\bar{v}_5) = [\bar{a}_3 \;\bar{b}_3\;\bar{c}_3]$, $(\bar{u}_2,\bar{v}_2) = [\bar{r}_1 \;\bar{s}_1\;\bar{w}_1]$, $(\bar{u}_2,\bar{v}_4) = [\bar{r}_2 \;\bar{s}_2\;\bar{w}_2]$, $(\bar{u}_2,\bar{v}_5) = [\bar{r}_3 \;\bar{s}_3\;\bar{w}_3]$, $(\bar{u}_3, \bar{v}_3) = [\hat{x}_1 \;\bar{y}_1\;\bar{z}_1]$, $(\bar{u}_3, \bar{v}_4) = [\hat{x}_2 \;\bar{y}_2\;\bar{z}_2]$, $(\bar{u}_3, \bar{v}_5) = [\hat{x}_3 \;\bar{y}_3\;\bar{z}_3]$, $\bar{e}_1 = \bar{a}$, for $2\leq i\leq q^\prime +3$: $e_i = \sum_{j=2, j \neq i}^{q^\prime +2} \bar{x}_i$. It can be easily seen that from these vectors all terminals can retrieve all of their demands.  
%
\end{IEEEproof}
\textit{Intuition}: The $\mathcal{M}_3$ sub-network of $\mathcal{N}_2$ has a $3$-dimensional VLNC solution over all finite fields (proved in \cite{das}). Hence, the terminal $\bar{t}_{25}$ needs no information from $\bar{e}_1$ of the Char-$q^\prime$-$\bar{x}$ network for it to compute its demands. The Char-$q^\prime$-$\bar{x}$ sub-network of $\mathcal{N}_2$ has an SLNC solution over all finite fields if the vector carried by $\bar{e}_1$ is a linear function of only $\bar{a}$ (Lemma~\ref{Junelemma1}).  Since both of these two sub-networks have a $3$-dimensional VLNC solution over all finite fields, $\mathcal{N}_2$ has a $3$-dimensional VLNC solution over all finite fields. 

\begin{lemma}\label{4Mlemma4}
$P(\mathcal{N}_2,5) = \{p\in \mathbb{P}| p \text{ divides } q^\prime \}$.
\end{lemma}
The proof is deferred to Appendix~\ref{app2.3}.




\section{Main Results}\label{sec3}
Let $\mathcal{G}_1$ and $\mathcal{G}_2$ be two arbitrary networks. Let $V_1$ be the set of all nodes of $\mathcal{G}_1$ and $E_1$ be the set of all edges of $\mathcal{G}_1$; let $V_2$ be the set of all nodes of $\mathcal{G}_2$ and $E_2$ be the set of all edges of $\mathcal{G}_2$.
\begin{definition}\label{union}
The union of networks $\mathcal{G}_1$ and $\mathcal{G}_2$ is denoted by $\mathcal{G}_1 \cup \mathcal{G}_2$, and the node set of $\mathcal{G}_1 \cup \mathcal{G}_2$ is $V_1 \cup V_2$, and edge set of $\mathcal{G}_1 \cup \mathcal{G}_2$ is $E_1 \cup E_2$.
\end{definition}
\begin{lemma}\label{Julylemma5}
Given $V_1 \cap V_2 = \emptyset$ and $E_1 \cap E_2 = \emptyset$, for any positive integer $d$, $P(\mathcal{G}_1 \cup \mathcal{G}_2,d) = P(\mathcal{G}_{1},d) \cap P(\mathcal{G}_{2},d)$.
\end{lemma}
\begin{IEEEproof}
If $\mathcal{G}_1$ and $\mathcal{G}_2$ both have a $d$-dimensional VLNC solution over a finite field, then it is immediate that $\mathcal{G}_1 \cup \mathcal{G}_2$ also has a $d$-dimensional VLNC solution over the same finite field; so $P(\mathcal{G}_1 \cup \mathcal{G}_2) \supseteq P(\mathcal{G}_{1},d) \cap P(\mathcal{G}_{2},d)$.  If $\mathcal{G}_1 \cup \mathcal{G}_2$ has a $d$-dimensional VLNC solution over a finite field, since there is no information exchange between $\mathcal{G}_1$ and $\mathcal{G}_2$ (as $\{V_1 \cup E_1\} \cap \{V_2 \cup E_2\} = \emptyset$), both of the sub-networks $\mathcal{G}_1$ and $\mathcal{G}_2$ have a $d$-dimensional VLNC solution; so $P(\mathcal{G}_1 \cup \mathcal{G}_2) \subseteq P(\mathcal{G}_{1},d) \cap P(\mathcal{G}_{2},d)$.
\end{IEEEproof}
%
%
\begin{theorem}\label{main:1}
For any finite set of primes $P = \{p_1,p_2,\ldots ,p_l\}$, there exists a network $\mathcal{N}_3$ for which $P(\mathcal{N}_3,1) = P$, and $P(\mathcal{N}_3,2) = \mathbb{P}$.
\end{theorem}
\begin{IEEEproof}
The network $\mathcal{N}_1$ (shown in Section~\ref{n1}) for $q = p_1\times p_2\times\cdots \times p_l$ satisfies the properties of $\mathcal{N}_3$ proposed in this theorem. Lemma~\ref{4Mlemma1} shows that $P(\mathcal{N}_1,1) = \{p \in \mathbb{P} | p \text{ divides } q\}$. For our chosen value of $q$, $p$ divides $q$ if and only if $p \in \{p_1,p_2,\ldots ,p_l\}$. Lemma~\ref{Julylemma1} shows that $P(\mathcal{N}_1,2) = \mathbb{P}$.
\end{IEEEproof}
\begin{theorem}\label{main:1a}
For any two non-empty sets of primes $P_1 = \{p_1,p_2,\ldots ,p_{l_1}\}$ and $P_2 = \{p_1^\prime,p_2^\prime,\ldots ,p_{l_2}^\prime\}$, there exists a network $\mathcal{N}_4$ such that $P(\mathcal{N}_4,1) = P_1$, but $P(\mathcal{N}_4,2) = \{P_1,P_2 \}$.
\end{theorem}
\begin{IEEEproof}
Consider the network $\mathcal{N}_1$ for $q = p_1\times p_2\times \cdots \times p_{l_1}$, and the network Char-$m$ for $m = p_1\times p_2\times \cdots \times p_{l_1}\times p_1^\prime\times p_2^\prime\times\cdots\times p_{l_2}^\prime$. Let the latter network be denoted by $\mathcal{C}$. Consider the union of the these two networks and let it be denoted by $\mathcal{C}\cup\mathcal{N}_1$. We show that the theorem holds for $\mathcal{N}_4 = \mathcal{C}\cup\mathcal{N}_1$. 

As argued in the earlier theorem (Theorem~\ref{main:1}), $P(\mathcal{N}_1,1) = P_1$. But as per Lemma~\ref{Julylemma2}, and our selected value of $m$, for any $d \in \mathbb{Z}^+$,  $P(\text{Char-}m,d) = \{P_1,P_2\}$. Then as per Lemma~\ref{Julylemma5}, $P(\mathcal{C}\cup\mathcal{N}_1,1) = P_1$. Lemma~\ref{Julylemma1} shows $P(\mathcal{N}_1,2) = \mathbb{P}$. Hence, again as per Lemma~\ref{Julylemma5}, $P(\mathcal{C}\cup\mathcal{N}_1,2) = \{P_1,P_2 \}$.
\end{IEEEproof}
\begin{theorem}\label{main:2}
For any two non-empty sets of primes $P_1 = \{p_1,p_2,\ldots ,p_{l_1}\}$ and $P_2 = \{p_1^\prime,p_2^\prime,\ldots ,p_{l_2}^\prime\}$, there exists a network $\mathcal{N}_5$ such that $P(\mathcal{N}_5,2) = P_1$, but $P(\mathcal{N}_5,3) = \{P_1,P_2 \}$.
\end{theorem}
\begin{IEEEproof}
Consider the network $\mathcal{N}_2$ for $q^\prime = p_1\times p_2\times \cdots \times p_{l_1}$, and the network Char-$m$ for $m = p_1\times p_2\times \cdots \times p_{l_1}\times p_1^\prime\times p_2^\prime\times\cdots\times p_{l_2}^\prime$. Let the latter network be denoted by $\mathcal{C}$. Consider the union of the these two networks and let it be denoted by $\mathcal{C}\cup\mathcal{N}_2$. We show that the theorem holds for $\mathcal{N}_5 = \mathcal{C}\cup\mathcal{N}_2$. 


From Lemma~\ref{4Mlemma3}, due to the selected value of $q^\prime$, we have $P(\mathcal{N}_2,2) = P_1$. But as per Lemma~\ref{Julylemma2}, and our selected value of $m$, for any $d \in \mathbb{Z}^+$,  $P(\text{Char-}m,d) = \{P_1,P_2\}$. Then as per Lemma~\ref{Julylemma5}, $P(\mathcal{C}\cup\mathcal{N}_2,2) = P_1$. Lemma~\ref{Julylemma6} shows $P(\mathcal{N}_2,3) = \mathbb{P}$. Hence, again as per Lemma~\ref{Julylemma5}, $P(\mathcal{C}\cup\mathcal{N}_2,3) = \{P_1,P_2\}$.
\end{IEEEproof}
%
\begin{remark}
Choosing $P_1$ and $P_2$ appropriately, $|P(\mathcal{N}_5,3)| - |P(\mathcal{N}_5,2)|$ can be made as large as wished.
\end{remark}
%
%
%
%
\begin{theorem}\label{main:4}
For any two non-empty sets of primes $P_1 = \{p_1,p_2,\ldots ,p_{l_1}\}$ and $P_2 = \{p_1^\prime,p_2^\prime,\ldots ,p_{l_2}^\prime\}$, there exists a network $\mathcal{N}_6$ such that $P(\mathcal{N}_6,2) = \{P_1,P_2 \}$, but $P(\mathcal{N}_6,3) = P_2$.
\end{theorem}
\begin{IEEEproof}
Consider the union of the networks $\mathcal{N}_1$ for $q = p_1^\prime\times p_2^\prime\times\cdots\times p_{l_2}^\prime$ and the network $\mathcal{N}_2$ for $q^\prime = p_1\times p_2\times \cdots \times p_{l_1}\times p_1^\prime\times p_2^\prime\times\cdots\times p_{l_2}^\prime$, and let resultant network be denoted by $\mathcal{N}_1\cup\mathcal{N}_2$. We show that the theorem holds for $\mathcal{N}_6 = \mathcal{N}_1\cup\mathcal{N}_2$. 

Lemma~\ref{4Mlemma2} shows that $P(\mathcal{N}_2,1) = \emptyset$. So $\mathcal{N}_1\cup\mathcal{N}_2$ has no SLNC solution. Lemma~\ref{Julylemma1} shows  $P(\mathcal{N}_1,2) = \mathbb{P}$ and Lemma~\ref{4Mlemma3} shows, for our selected value of $q^\prime$, $P(\mathcal{N}_1,2) = \{P_1,P_2\}$. Hence, as per Lemma~\ref{Julylemma5}, $P(\mathcal{N}_1\cup\mathcal{N}_2,2) = \{P_1,P_2\}$. Lemma~\ref{Julylemma6} shows $P(\mathcal{N}_2,3) = \mathbb{P}$, and Lemma~\ref{4Mlemma1} shows, for our selected value of $q$, $P(\mathcal{N}_1,3) = P_2$. Hence, as per Lemma~\ref{Julylemma5}, $P(\mathcal{N}_1\cup\mathcal{N}_2,3) = P_2$.
\end{IEEEproof}
\begin{remark}
Choosing $P_1$ and $P_2$ appropriately, $|P(\mathcal{N}_6,2)| - |P(\mathcal{N}_6,3)|$ can be made as large as wished.
\end{remark}
\begin{theorem}\label{main:5}
There exists a network which has a $2$-dimensional VLNC solution and a $3$-dimensional VLNC solution, but has no $5$-dimensional VLNC solution.
\end{theorem}
\begin{IEEEproof}
Let $P_1 = \{p_1,p_2,\ldots ,p_{l_1}\}$ and $P_2 = \{p_1^\prime,p_2^\prime,\ldots ,p_{l_2}^\prime\}$ be two disjoint sets of primes (\textit{i.e.} $P_1 \cap P_2 = \emptyset$). Consider the union of the networks $\mathcal{N}_1$ for $q = p_1^\prime\times p_2^\prime\times\cdots\times p_{l_2}^\prime$ and the network $\mathcal{N}_2$ for $q^\prime = p_1\times p_2\times \cdots \times p_{l_1}$, and let resultant network be denoted by $\mathcal{N}_1\cup\mathcal{N}_2$. We show that the network $\mathcal{N}_1\cup\mathcal{N}_2$ satisfies the theorem. 

Similar to the proof of Theorem~\ref{main:4} it can be shown that $P(\mathcal{N}_1\cup\mathcal{N}_2,2) = P_1$, and $P(\mathcal{N}_1\cup\mathcal{N}_2,3) = P_2$.

Lemma~\ref{4Mlemma1} shows that $P(\mathcal{N}_1,5) = P_2$, and Lemma~\ref{4Mlemma4} shows that for our selected value of $q^\prime$, $P(\mathcal{N}_2,5) = P_1$. But $P_1 \cap P_2 = \emptyset$. Hence, $P(\mathcal{N}_1\cup\mathcal{N}_2,5) = \emptyset$. This proves the theorem. 
\end{IEEEproof}
Theorem~\ref{main:5} proves the following Corollary.
\begin{corollary}
A network having an $m_1$-dimensional VLNC solution and an $m_2$-dimensional VLNC solution may not have a $(m_1 + m_2)$-dimensional VLNC solution. 
\end{corollary}

%
%
%
%
\begin{figure*}
\centering
\includegraphics[width=0.6\textwidth]{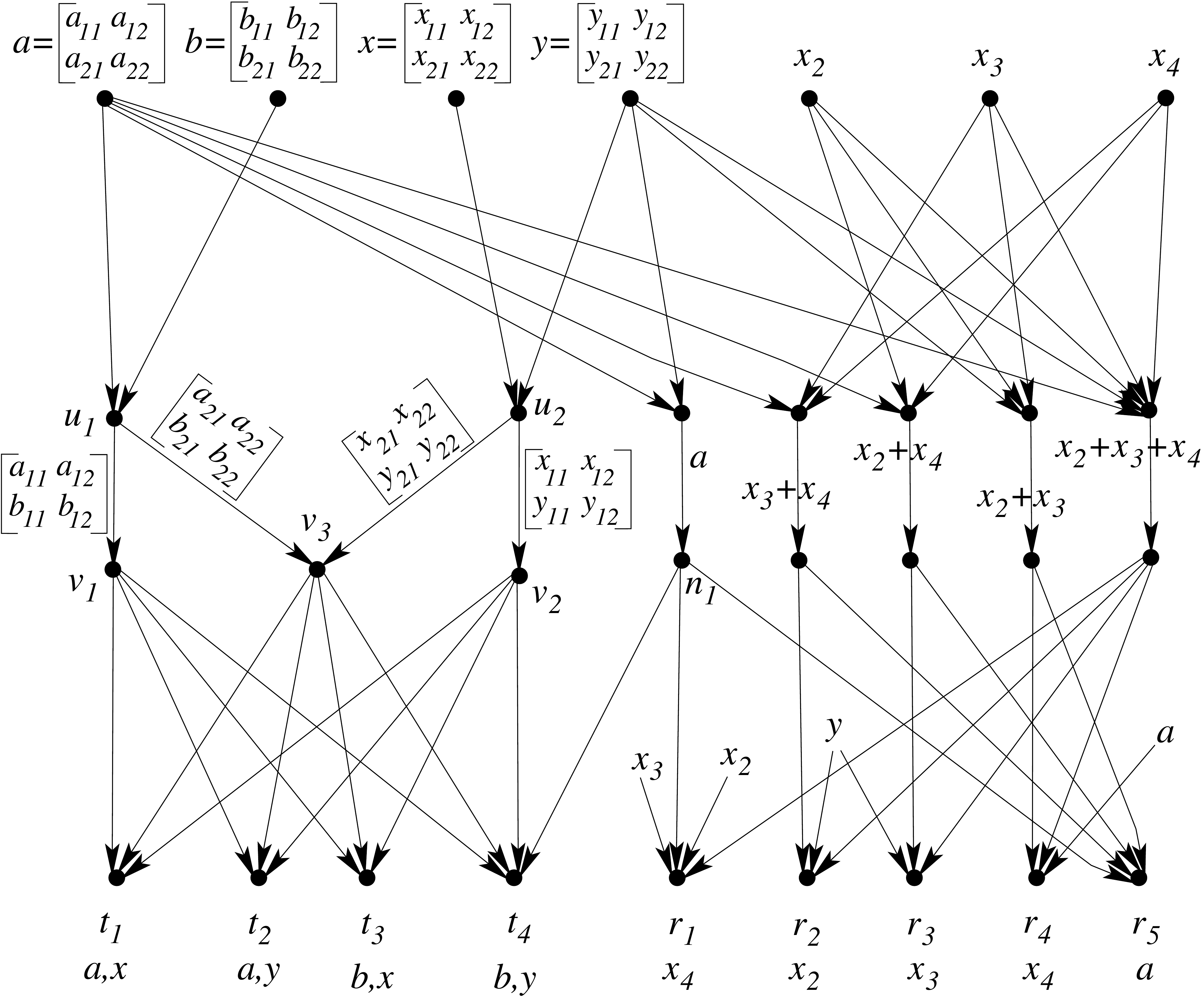}
\caption{An SLNC solution of the network $\mathcal{N}_1$ for $q = 2$ over a non-commutative ring of $2\times 2$ square matrices. The codes for the M-network (sub-network of $\mathcal{N}_1$) is taken from \cite{connelly2}.}
\label{codes2}
\end{figure*}
\begin{theorem}\label{main:7}
For any prime number $p$, there exists a network which has an SLNC solution over a finite field if and only if the size of the finite field is a positive integer power of $p$, but has an SLNC solution over a non-commutative ring of size $16$.
\end{theorem}
\begin{IEEEproof}
We show that the network $\mathcal{N}_1$ for $q = p$ is such a network. Lemma~\ref{4Mlemma1} shows that $\mathcal{N}_1$ has an SLNC solution over a finite field if and only if the characteristic of the finite field divides $p$, which can only happen if the characteristic is $p$.

In \cite{connelly2}, the authors showed that the M-network has an SLNC solution over a non-commutative ring of size $16$. On the other hand, from the proof of Lemma~\ref{Junelemma1}, it can be seen that only addition and subtraction operations are required to achieve an SLNC solution of the Char-$q$-$y$ sub-network (the source $y$ is labelled as $s$ in the proof of Lemma~\ref{Junelemma1}) of $\mathcal{N}_1$ when $e_1$ is a linear function of only $a$ and not of both $a$ and $y$. Hence, the same solution would also work over any ring. Intuitively it can be seen that since both of the constituent M-network and the Char-$q$-$y$ network have an SLNC solution over a non-commutative ring of size $16$, network $\mathcal{N}_1$ also has an SLNC solution over the same ring. A coding scheme is shown next.

The SLNC solution of the M-network over a non-commutative ring of size $16$, which was shown in \cite{connelly2}, is used in Fig.~\ref{codes2} to produce an SLNC solution for the network $\mathcal{N}_1$ for $q = 2$. Let the source alphabet be the non-commutative ring of $2\times 2$ square matrices over $\mathbb{F}_2$. Not shown in the figure: edge $(v_3,t_1)$ carries $\begin{bmatrix} a_{21} & a_{22} \\ x_{21} & x_{22} \end{bmatrix}$, edge $(v_3,t_2)$ carries $\begin{bmatrix} a_{21} & a_{22} \\ y_{21} & y_{22} \end{bmatrix}$, edge $(v_3,t_3)$ carries $\begin{bmatrix} b_{21} & b_{22} \\ x_{21} & x_{22} \end{bmatrix}$, and edge $(v_3,t_4)$ carries $\begin{bmatrix} b_{21} & b_{22} \\ y_{21} & y_{22} \end{bmatrix}$.  Terminal $t_1$ retrieves the demands using the following operations: $a = \begin{bmatrix} 1 & 0 \\ 0 & 0 \end{bmatrix} \begin{bmatrix} a_{11} & a_{12} \\ b_{11} & b_{12} \end{bmatrix} + \begin{bmatrix} 0 & 0 \\ 1 & 0 \end{bmatrix}  \begin{bmatrix} a_{21} & a_{22} \\ x_{21} & x_{22} \end{bmatrix} $, and $x = \begin{bmatrix} 1 & 0 \\ 0 & 0 \end{bmatrix} \begin{bmatrix} x_{11} & x_{12} \\ y_{11} & y_{12} \end{bmatrix} + \begin{bmatrix} 0 & 0 \\ 0 & 1 \end{bmatrix}  \begin{bmatrix} a_{21} & a_{22} \\ x_{21} & x_{22} \end{bmatrix}$. Terminals $t_2,t_3$, and $t_4$ can retrieve their demands similarly. The decoding for the terminals $r_i$ is similar to that shown in Lemma~\ref{Junelemma1}. This same SLNC solution can be easily extended for other value of $q$.  
\end{IEEEproof}
\begin{remark}
The network $\mathcal{N}_1$ for an appropriate value of $q$, has the property that the least sized finite field over which it has an SLNC solution is arbitrarily larger than the least sized non-commutative ring over which it has an SLNC solution.
%
\end{remark}
\begin{remark}
It has been shown in \cite{connelly1} that the least sized commutative ring over which a network has an SLNC solution is a field. Hence, from Theorem~\ref{main:7} it can be concluded that for any $p>16$ there exists a network which admits an SLNC over a non-commutative ring whose size is less than any commutative ring over which the network admits an SLNC solution.
\end{remark}


\section{Conclusion}\label{sec4}
In this paper, we showed that the set of characteristics over which a VLNC solution exists depend upon the message dimension: as the message dimension is increased, the set of characteristics over which a VLNC solution exists may get larger or smaller. To the best of our knowledge, such a behaviour has never been reported earlier in the literature.

Recently it has been shown in \cite{LNCrings} that the linear coding capacity is dependent only on the characteristic of the finite field. Our results show that even if the linear coding capacity is achievable if all finite fields are considered, to achieve the linear coding capacity, one has to operate over appropriate finite fields depending on the chosen message dimension. Our work has made the importance of message dimension explicit, \textit{i.e.}, both the operational finite field and message dimension have to chosen appropriately to achieve the linear coding capacity.


We also showed that (non-commutative) rings are superior to finite fields in terms of achieving an SLNC solution over a lesser sized alphabet. But, whether rings are also superior when the objective is to achieve a VLNC solution. That is, whether there exists a network which, for any integer $d > 1$, has a $d$-dimensional VLNC solution over a finite field only if the size of the finite field is greater than or equal to $n$, but the same network has $d$-dimensional VLNC solution over a ring whose size is strictly less than $n$. This problem remains open.

\appendices

\section{Discrete Polymatroids}
It has been shown that a network has a $d$-dimensional VLNC solution if and only if a discrete polymatroid with certain properties exists \cite{matroid, sundar}. We have used this connection to establish some of the theorems in this paper. For the reader's convenience, we reproduce some definitions and theorems from \cite{sundar} that describes this connection.

Define $G = \{1,2,\ldots n\}$, $\mathbb{Z}_{\geq 0}$ as the set of non-negative integers, and $\mathbb{Z}_{\geq 0}^{n}$ as the set of all $n$ length vectors over $\mathbb{Z}_{\geq 0}$. For a vector $v$ and a set $A \subseteq G$, let $v(A)$ be the vector having only the components indexed by the elements of $A$, and $|v(A)|$ denote the sum (over integers) of the components of $v(A)$. For example, if $v = (0,2,1,0)$ and $A = \{1,3\}$, then $v(A) = (0,1)$, and $|v(A)| = 1$.

\begin{definition}[\textit{Defintion 2}, \cite{sundar}]\label{P}
Let $\rho: 2^G \rightarrow \mathbb{Z}_{\geq 0}$ such that
\begin{description}
\item[{[P1]}] $\rho(\emptyset) = 0$
\item[{[P2]}] $\rho(A) \leq \rho(B)$ if $A \subseteq B$
\item[{[P3]}] $\rho(A) + \rho(B) \geq \rho(A\cup B) + \rho(A\cap B)$
\end{description}
Let $\mathbb{D} = \{x\in \mathbb{Z}_{\geq 0}^{n} | \text{ such that } |x(A)| \leq \rho(A), \forall A\subseteq G\}$. Then $\mathbb{D}$ is a discrete polymatroid with rank function $\rho$ and ground set $G$.
\end{definition} 
\begin{example}\label{ex1}
\normalfont Let $G = \{1,2,3\}$ and $\rho(\emptyset) = 0$, $\rho(\{1\}) = \rho(\{2\}) = \rho(\{1,2\}) = 1$, $\rho(\{3\}) = \rho(\{1,3\}) = 2$, $\rho(\{2,3\}) = \rho(\{1,2,3\}) = 3$. It can be seen that $\rho$ does not follow [P3] of Definition~\ref{P} as $\rho(\{1,2\}) + \rho(\{1,3\}) \geq \rho(\{1,2,3\}) + \rho(\{1\})$ returns $3 \geq 4$.
\end{example}
\begin{example}\label{ex2}
\normalfont In Example~\ref{ex1}, let $\rho(\{2,3\}) = \rho(\{1,2,3\}) = 2$. Then $\rho$  obeys conditions [P1]--[P3], and we have $\mathbb{D} = \{(0,0,0), (1,0,0), (0,1,0), (0,0,1), (0,0,2), (1,0,1), (0,1,1)\}$.
\end{example}
\begin{definition}[\textit{Defintion 3}, \cite{sundar}]\label{PR}
A discrete polymatroid $\mathbb{D}$ with rank function $\rho$ and ground set $G$ is said to be representable over $\mathbb{F}_q$ if for each element $i$ of $G$, there exists a vector subspace $V_i$ of a vector space $V$  over $\mathbb{F}_q$ such that $dim(\sum_{i\in X} V_i) = \rho(X)$ for $\forall X\subseteq G$.
\end{definition}
\begin{example}
\normalfont In Example~\ref{ex2}, the polymatroid $\mathbb{D}$ is representable over all finite fields. If $v$ is a vector, then let $<v>$ denote the vector space spanned by $v$. Now, let $V$ be the vector space $\mathbb{F}_q^{\,2}$, with subspaces $V_1 = <(1,0)>$, $V_2 = <(1,0)>$, $V_3 = <(1,0),(0,1)>$. It can be seen that $V_1,V_2,V_3$ forms a representation of $\mathbb{D}$. 
\end{example}
%
%
Let $\epsilon_{in}$ be an $n$ length vector whose $i^{th}$ component is one and all other components are zero. If $s$ is a source, then define $In(s) = s$, and $Out(s)$ as the set of all edges whose tail node is $s$.  If $v$ is an intermediate note, then define $In(v)$  as the set of edges whose head node is $v$, and $Out(v)$ as the set of edges whose tail node is $v$. If $t$ is a terminal, then define $In(t)$ as the set of all edges whose head node is $t$, and $Out(v)$ as the set of all sources demanded by $t$. The following theorem combines \textit{Definition~7} and \textit{Theorem~1} of \cite{sundar}.
\begin{theorem}\label{DP}[\textit{Defintion 7} and \textit{Theorem 1}, \cite{sundar}]
For a network $\mathcal{N}$ let the set of sources be $S$, the set of non-source nodes be $V$, and the set of edges be $E$. The network $\mathcal{N}$ has a $d$-dimensional VLNC solution over $\mathbb{F}_q$ if and only if there exists a discrete polymatroid $\mathbb{D}$ with rank function $\rho$, and ground set $G$, such that $\mathbb{D}$ representable over $\mathbb{F}_q$, and there exists a map $f:\{S\cup E\} \rightarrow G$ that satisfies the following conditions:
\begin{description}
\item[{[D1]}] $f$ is one-to-one on $S$.
\item[{[D2]}] $\sum_{i\in f(S)} d\epsilon_{in}\in\mathbb{D}$.
\item[{[D3]}] $\forall s\in S$, $\rho(f(s)) = d$, and $\forall e\in E$, $\rho(f(e)) \leq d$.
\item[{[D4]}] $\rho(f(In(v))) = \rho(f(In(v)\cup Out(v)))$, $\forall v \in V$. 
\end{description}
\end{theorem}
Reference \cite{sundar} also shows that a $d$-dimensional VLNC solution of $\mathcal{N}$ can be constructed from the representation of $\mathbb{D}$. The condition [D1] indicates that the sources are mapped to separate elements of the ground set (also, each source corresponds to a separate vector subspace in the representation). Condition [D2] captures the notion that the uncertainty associated with the messages generated by one source cannot be reduced by knowing the messages generated by the other sources (also, the corresponding vector subspaces in the representation are mutually disjoint). [D3] indicates that each source generates $d$ symbols, and each edge carries a maximum of $d$ symbols. [D4] captures the fact that the symbols carried by the edges outgoing from a node or decoded by a node is a function of the symbols generated by the node or carried by the edges incoming to the node.  

We now prove two lemmas (Lemma~\ref{toughlemma} and Lemma~\ref{lema}) which are part of the existing literature, but to our knowledge, have never been explicitly shown.

For a network having a $d$-dimensional VLNC solution, let $S$ be its set of sources, and let $\mathbb{D}$ be the corresponding discrete polymatroid whose existence is guaranteed by Theorem~\ref{DP}. Say $S_1$ and $S_2$ are two subsets of $S$. Let $f$ be the function that maps the sources and edges of the network to the ground set of $\mathbb{D}$ conforming to Theorem~\ref{DP}, and $\rho$ be the rank function of $\mathbb{D}$. Define $\mathtt{g} = \rho \circ f$.  
\begin{lemma}\label{toughlemma}
$\mathtt{g}(S_1,S_2) = \mathtt{g}(S_1) + \mathtt{g}(S_2)$.
\end{lemma}
\begin{IEEEproof}
For simplicity, we prove for a particular case when $S_1 = \{s_1,s_2\}$ and $S_2=\{s_3,s_4\}$; other possibilities can be proved similarly. Note that according to [D1] of Theorem~\ref{DP}, all sources are mapped to different elements. Now, if the ground set of $\mathbb{D}$ is $\{1,2,\ldots,n\}$, then according to [D2] of Theorem~\ref{DP},  the vector $v = \sum_{i \in f(S)} d\epsilon_{in}$ is in $\mathbb{D}$. Hence, from Definition~\ref{P}, the vector $d\epsilon_{f(s_1)n} + d\epsilon_{f(s_2)n} + d\epsilon_{f(s_3)n} + d\epsilon_{f(s_4)n}$ is in $\mathbb{D}$. So,  $4d \leq \rho(\{f(s_1),f(s_2),f(s_3),f(s_4)\})$. Also, from [D3] of Theorem~\ref{DP}, we have: $\rho(f(s_1)) = \rho(f(s_2)) = \rho(f(s_3)) = \rho(f(s_4)) = d$. So,
\begin{equation}
\rho(f(s_1)) + \rho(f(s_2)) + \rho(f(s_3)) + \rho(f(s_4)) \leq \rho(\{f(s_1),f(s_2),f(s_3),f(s_4)\})\label{lab1}
\end{equation}
On the other hand, from [P3] of Definition~\ref{P}, we have:
\begin{equation}
\rho(\{f(s_1),f(s_2),f(s_3),f(s_4)\}) \leq \rho(f(s_1)) + \rho(f(s_2)) + \rho(f(s_3)) + \rho(f(s_4))\label{lab2}
\end{equation}
From equations (\ref{lab1}) and (\ref{lab2}), we must have: $\rho(\{f(s_1),f(s_2),f(s_3),f(s_4)\}) = \rho(f(s_1)) + \rho(f(s_2)) + \rho(f(s_3)) + \rho(f(s_4))$.
\end{IEEEproof}
%
\begin{lemma}\label{lema1}
If $C\subseteq B$, then $\mathtt{g}(A,B) - \mathtt{g}(A,C) \leq \mathtt{g}(B) - \mathtt{g}(C)$.
\end{lemma}
\begin{IEEEproof}
\begin{IEEEeqnarray*}{l}
\mathtt{g}(A,C) + \mathtt{g}(B) \geq \mathtt{g}(A,B,C) + \mathtt{g}(C) \qquad \text{[from [P3] of Definition~\ref{P}]}\\
\text{or, } \mathtt{g}(A,C) + \mathtt{g}(B) \geq \mathtt{g}(A,B) + \mathtt{g}(C) \\
\text{or, } \mathtt{g}(A,B) - \mathtt{g}(A,C) \leq \mathtt{g}(B) - \mathtt{g}(C) 
\end{IEEEeqnarray*}
\end{IEEEproof}
\begin{lemma}\label{lema}
For a network, let $S_1$ and $S_2$ be two subsets of the set of sources, and $E_1$ and $E_2$ be two subsets of the set of edges, such that $\mathtt{g}(S_1,E_1) = \mathtt{g}(S_1)$ and $\mathtt{g}(S_2,E_2) = \mathtt{g}(S_2)$. Then, if $\bar{S}_1$ is a subset of $S_1$ and $\bar{S}_2$ is a subset of $S_2$, $\mathtt{g}(\bar{S}_1,E_1) + \mathtt{g}(\bar{S}_2,E_2) = \mathtt{g}(\bar{S}_1,E_1,\bar{S}_2,E_2)$.
\end{lemma}
\begin{IEEEproof}
Due to [P3] of Definition~\ref{P}, Lemma~\ref{toughlemma}, and the proposition of this lemma, we have:
\begin{equation}
\mathtt{g}(S_1,E_1,S_2,E_2) \leq \mathtt{g}(S_1,E_1) + \mathtt{g}(S_2,E_2) =  \mathtt{g}(S_1) + \mathtt{g}(S_2) = \mathtt{g}(S_1,S_2) \leq \mathtt{g}(S_1,E_1,S_2,E_2)
\end{equation}
So we must have: 
\begin{equation}
\mathtt{g}(S_1,E_1) + \mathtt{g}(S_2,E_2) = \mathtt{g}(S_1) + \mathtt{g}(S_2) = \mathtt{g}(S_1,S_2) = \mathtt{g}(S_1,E_1,S_2,E_2)\label{hhh2}
\end{equation}
Then,
\begin{IEEEeqnarray*}{l}
\mathtt{g}(S_2,E_2) - \mathtt{g}(\bar{S}_2,E_2)\\
= \mathtt{g}(S_1,E_1,S_2,E_2) - \mathtt{g}(S_1,E_1) - \mathtt{g}(\bar{S}_2,E_2) \qquad \text{ [using equation~(\ref{hhh2})]}\\
\leq \mathtt{g}(S_1,E_1,S_2,E_2) - \mathtt{g}(S_1,E_1,\bar{S}_2,E_2)\qquad \text{ [applying [P3] of Definition~\ref{P}]} \IEEEyesnumber\label{oo99}\\
\leq \mathtt{g}(S_2,E_2) - \mathtt{g}(\bar{S}_2,E_2)  \qquad \text{ [taking $A= \{S_1\cup E_1\}$ in Lemma~\ref{lema1}]} \IEEEyesnumber\label{oo1}
\end{IEEEeqnarray*}
From equations~(\ref{oo99}) and (\ref{oo1}), we have:
\begin{equation}
\mathtt{g}(S_1,E_1,S_2,E_2) - \mathtt{g}(S_1,E_1,\bar{S}_2,E_2) = \mathtt{g}(S_2,E_2) - \mathtt{g}(\bar{S}_2,E_2)\label{oo2}
\end{equation}
Then,
\begin{IEEEeqnarray*}{l}
\mathtt{g}(S_1,E_1) - \mathtt{g}(\bar{S}_1,E_1)\\
= \mathtt{g}(S_1,E_1,S_2,E_2) - \mathtt{g}(S_2,E_2) - \mathtt{g}(\bar{S}_1,E_1)\qquad \text{ [using equation~(\ref{hhh2})]}\\
\leq \mathtt{g}(S_1,E_1,S_2,E_2) - \mathtt{g}(\bar{S}_1,E_1,S_2,E_2)\qquad \text{ [applying [P3] of Definition~\ref{P}]}\\
\leq \mathtt{g}(S_1,E_1,\bar{S}_2,E_2) - \mathtt{g}(\bar{S}_1,E_1,\bar{S}_2,E_2)\qquad  \text{ [taking $A = S_2\setminus \bar{S}_2$ in Lemma~\ref{lema1}]}\IEEEyesnumber\label{oo98}\\
\leq \mathtt{g}(S_1,E_1) - \mathtt{g}(\bar{S}_1,E_1) \qquad \text{ [taking $A = \bar{S}_2\cup E_2$ in Lemma~\ref{lema1}]}\IEEEyesnumber\label{oo3}
\end{IEEEeqnarray*}

From equations~(\ref{oo98}) and (\ref{oo3}), we have:
\begin{equation}
\mathtt{g}(S_1,E_1,\bar{S}_2,E_2) - \mathtt{g}(\bar{S}_1,E_1,\bar{S}_2,E_2)  = \mathtt{g}(S_1,E_1) - \mathtt{g}(\bar{S}_1,E_1)\label{oo4}
\end{equation}

Adding equations~(\ref{oo2}) and (\ref{oo4}), we get:
\begin{IEEEeqnarray*}{l}
\mathtt{g}(S_1,E_1,S_2,E_2) - \mathtt{g}(S_1,E_1,\bar{S}_2,E_2) + \mathtt{g}(S_1,E_1,\bar{S}_2,E_2) - \mathtt{g}(\bar{S}_1,E_1,\bar{S}_2,E_2) \\\hfill =\> \mathtt{g}(S_2,E_2) - \mathtt{g}(\bar{S}_2,E_2) + \mathtt{g}(S_1,E_1) - \mathtt{g}(\bar{S}_1,E_1)\\
\text{or, } \mathtt{g}(S_1,E_1,S_2,E_2) - \mathtt{g}(\bar{S}_1,E_1,\bar{S}_2,E_2) = \mathtt{g}(S_1,E_1,S_2,E_2) - \mathtt{g}(\bar{S}_2,E_2) - \mathtt{g}(\bar{S}_1,E_1)\qquad \text{ [using equation~\ref{hhh2}]}\\
\text{or, } \mathtt{g}(\bar{S}_1,E_1,\bar{S}_2,E_2) = \mathtt{g}(\bar{S}_2,E_2) + \mathtt{g}(\bar{S}_1,E_1)
\end{IEEEeqnarray*}
\end{IEEEproof}
Consider the  Char-$q$-$s$ network shown in Section~\ref{char} and let $p$ be a prime such that $p$ does not divide $q$. Let $f$ be the function that maps the sources and edges of the  Char-$q$-$s$ network to the ground set $G$ of a discrete polymatrid $\mathbb{D}$ with rank function $\rho$ such that Char-$q$-$s$ network has a $d$-dimensional VLNC solution over a finite field $\mathbb{F}_{p^n}$ ($n$ is positive integer) if and only if $\mathbb{D}$ is representable over $\mathbb{F}_{p^n}$ and $f$ follows the conditions given in Theorem~\ref{DP}. Now let $\mathtt{g} = \rho \circ f$.
\begin{lemma}\label{july63}
For the Char-$q$-$s$ network, if $p$ does not divide $q$, then over a finite field $\mathbb{F}_{p^n}$, $\mathtt{g}(x_1) = \mathtt{g}(x_1,e_1)$.
\end{lemma}
\begin{IEEEproof}
In Lemma~\ref{Junelemma2}, it can be seen that when the characteristic of the finite field divides $q$, (i) $A_1 = 0$, and (ii) $Z_1M_1 = I$ (equation~(\ref{10f34})). So by the operation $Z_1y_{e_1}$, the source vector $x_1$ can be retrieved from $y_{e_1}$. This shows that the following equation is true.
\begin{equation}
\mathtt{g}(e_1) = \mathtt{g}(e_1,x_1) \label{july61}
\end{equation}
From [P2] of Definition~\ref{P}, we know that: 
\begin{equation}
\mathtt{g}(x_1) \leq \mathtt{g}(x_1,e_1) \label{july62}
\end{equation}
Substituting equation~(\ref{july61}) in equation~(\ref{july62}), we get: $\mathtt{g}(x_1) \leq \mathtt{g}(e_1)$. On the other hand, [D3] of Theorem~\ref{DP} shows that since $x_1$ is a source and $e_1$ is an edge, $\mathtt{g}(x_1) \geq \mathtt{g}(e_1)$. So we must have: $\mathtt{g}(x_1) = \mathtt{g}(e_1)$. Substituting this result in equation~(\ref{july61}), we get: $\mathtt{g}(x_1) = \mathtt{g}(e_1,x_1)$. 
\end{IEEEproof}
%
%
%

%
%
%
%
\section{Proof of Lemma~\ref{4Mlemma1}}\label{app1}
\begin{IEEEproof}
Let $f$ be the function that maps the network $\mathcal{N}_1$ to a discrete polymatroid $\mathbb{D}_1$ conforming to the conditions given in Theorem~\ref{DP}. Let $\rho$ be the rank function of $\mathbb{D}_1$, and let $\mathtt{g} = \rho \circ f$. 
Consider the `only if' part. We show that if the characteristic of the finite field does not divide $q$, then $\mathcal{N}_1$ has no odd dimensional VLNC solution. Let us assume that over a finite field whose characteristic does not divide $q$, $\mathcal{N}_1$ has a $d$-dimensional VLNC solution for some odd positive integer $d$. 
Let the edges $(u_i,v_j)$ be denoted by $e_{ij}$ for $i=1,2$ and $j=1,2,3$. 
From [D4] of Theorem~\ref{DP}, we have $\mathtt{g}(a,b) = \mathtt{g}(a,b,e_{11})$ and $\mathtt{g}(x,y) = \mathtt{g}(x,y,e_{22})$. From Lemma~\ref{toughlemma}, we also have $\mathtt{g}(a,b) + \mathtt{g}(x,y) = \mathtt{g}(a,b,x,y)$. So this means $\mathtt{g}(a,b,e_{11}) + \mathtt{g}(x,y,e_{22}) = \mathtt{g}(a,b,e_{11},x,y,e_{22})$. Then, due to the demands of terminal $t_{1}$, we get the following.
\begin{IEEEeqnarray*}{l}
\mathtt{g}(e_{11},a) + \mathtt{g}(e_{22},x)\\
= \mathtt{g}(e_{11},a,e_{22},x) \qquad \text{ [using Lemma~\ref{lema}]}\\
\leq \mathtt{g}(e_{11},a,e_{22},x,(v_3,t_1))\\
= \mathtt{g}(e_{11},e_{22},(v_3,t_1)) \qquad \text{ [due to demands of $t_1$]}\\
\leq  \mathtt{g}(e_{11}) + \mathtt{g}(e_{22}) + \mathtt{g}((v_3,t_1)) 
\leq 3d \qquad \text{ [using [P3] of Definiton~\ref{P} and [D3] of Theorem~\ref{DP}]}\IEEEyesnumber\label{21f1}
\end{IEEEeqnarray*}
Similar to equation (\ref{21f1}), due to the demands of $t_2$ and $t_3$, we have the following equations.
\begin{IEEEeqnarray}{l}
\mathtt{g}(e_{11},a) + \mathtt{g}(e_{22},y) \leq 3d \label{21f2}\\
\mathtt{g}(e_{11},b) + \mathtt{g}(e_{22},x) \leq 3d \label{21f3}
\end{IEEEeqnarray}
Since the characteristic of the finite field does not divide $q$, from Lemma~\ref{Junelemma2}, we know that the message carried by $e_1$ is a linear function of only $a$.
\begin{IEEEeqnarray*}{l}
\mathtt{g}(e_{11},a) + d + d
\geq \mathtt{g}(e_{11},a) + \mathtt{g}(e_{22}) + \mathtt{g}((v_3,t_4))\\
\geq \mathtt{g}(e_{11},a,e_{22},(v_3,t_4))\\
= \mathtt{g}(e_{11},a,e_{22},(v_3,t_4),e_1) \qquad \text{ [since $\mathtt{g}(a) = \mathtt{g}(a,e_1)$ by Lemma~\ref{july63}]}\\
= \mathtt{g}(e_{11},a,e_{22},(v_3,t_4),e_1,b,y) \qquad \text{ [due to demands of $t_4$]}\\
\geq \mathtt{g}(e_{11},a,e_{22},b,y)\\
= \mathtt{g}(e_{11},a,b) + \mathtt{g}(e_{22},y) \qquad \text{ [from Lemma~\ref{lema}]}\\
= 2d + \mathtt{g}(e_{22},y) \IEEEyesnumber\label{21f4}
\end{IEEEeqnarray*}
From equation (\ref{21f4}), we get that
\begin{equation}
\mathtt{g}(e_{11},a) \geq \mathtt{g}(e_{22},y)\label{21f11}
\end{equation}
We know:
\begin{IEEEeqnarray*}{l}
4d = \mathtt{g}(a,b,x,y)\\
= \mathtt{g}(a,b,x,y,e_{11},e_{13},e_{22},e_{23})\\
= \mathtt{g}(e_{11},e_{13},e_{22},e_{23})\\
\leq \mathtt{g}(e_{11}) + \mathtt{g}(e_{13}) + \mathtt{g}(e_{22}) + \mathtt{g}(e_{23})\\
\leq 4d \qquad\text{ [using [D3] of Theorem~\ref{DP}]} \IEEEyesnumber\label{21f5}
\end{IEEEeqnarray*}
From equation (\ref{21f5}), we get:
\begin{equation}
\mathtt{g}(e_{11}) = \mathtt{g}(e_{13}) = \mathtt{g}(e_{22}) = \mathtt{g}(e_{23}) = d\label{21f6}
\end{equation}
We also have:
\begin{IEEEeqnarray*}{l}
\mathtt{g}(e_{11},a) + \mathtt{g}(e_{11},b)\\
\geq \mathtt{g}(e_{11},a,b) + \mathtt{g}(e_{11}) \qquad\text{ [using [P3] of Definition~\ref{P}]}\\
= \mathtt{g}(a,b) + \mathtt{g}(e_{11})\\
= 3d \qquad \text{ [using equation (\ref{21f6})]}\IEEEyesnumber\label{21f7}
\end{IEEEeqnarray*}
Similar to equation (\ref{21f7}), we have:
\begin{equation}
\mathtt{g}(e_{22},x) + \mathtt{g}(e_{22},y) \geq 3d\label{21f8}
\end{equation}
Adding equations (\ref{21f1}) and (\ref{21f2}), we get:
\begin{IEEEeqnarray*}{l}
2\mathtt{g}(e_{11},a) + \mathtt{g}(e_{22},x) + \mathtt{g}(e_{22},y) \leq 6d\\
\text{or, }2\mathtt{g}(e_{11},a) \leq 3d \qquad \text{ [substituting equation (\ref{21f8})]}\\
\text{or, }\mathtt{g}(e_{11},a) \leq \frac{3d}{2}\IEEEyesnumber\label{21f9}
\end{IEEEeqnarray*}
Adding equations (\ref{21f1}) and (\ref{21f3}), we get:
\begin{IEEEeqnarray*}{l}
\mathtt{g}(e_{11},a) + \mathtt{g}(e_{11},b) + 2\mathtt{g}(e_{22},x) \leq 6d\\
\text{or, }2\mathtt{g}(e_{22},x) \leq 3d \qquad \text{ [substituting equation (\ref{21f7})]}\\
\text{or, }\mathtt{g}(e_{22},x) \leq \frac{3d}{2}\IEEEyesnumber\label{21f10}
\end{IEEEeqnarray*}
From equations (\ref{21f11}) and (\ref{21f9}), we have:
\begin{equation}
\mathtt{g}(e_{22},y) \leq \frac{3d}{2}\label{21f12}
\end{equation}
Since $d$ is an odd positive integer, let $d = 2n -1$ where $n$ is a positive integer. Then, from equations (\ref{21f10}) and (\ref{21f12}), we have:
\begin{IEEEeqnarray}{l}
\mathtt{g}(e_{22},x) \leq \frac{3(2n-1)}{2} = 3n - \frac{3}{2} = 3n - 2 + \frac{1}{2}\label{21f13}\\
\mathtt{g}(e_{22},y) \leq \frac{3(2n-1)}{2} = 3n - \frac{3}{2} = 3n - 2 + \frac{1}{2}\label{21f14}
\end{IEEEeqnarray}
Since the rank function $\mathtt{g}()$ is integer valued by Definition~\ref{P}, from equations (\ref{21f13}) and (\ref{21f14}), we have:
\begin{IEEEeqnarray}{l}
\mathtt{g}(e_{22},x) \leq 3n - 2 \label{21f15}\\
\mathtt{g}(e_{22},y) \leq 3n - 2 \label{21f16}
\end{IEEEeqnarray}
Substituting values from equation (\ref{21f15}) and (\ref{21f16}) in equation (\ref{21f8}), we get:
\begin{equation}
6n - 4 \geq 3d = 3(2n-1) = 6n -3\label{21f17}
\end{equation}
Equation (\ref{21f17}) results in $3 \geq 4$, which is a contradiction.

We now show the `if' part of the proof. We show that $\mathcal{N}_1$ has an SLNC solution if the characteristic of the finite field divides $q$. Let the message vector generated by a source be denoted by the same label as the source. In our solution, the edges $e_i$ for $1\leq i\leq q+3$ carry the messages as indicated by equations (\ref{23f1})-(\ref{23f6}) with $x_1$ replaced by $a$ and $s$ replaced by $y$. Then, the terminals $r_1,r_2,\ldots ,r_{q+3}$ can retrieve its desired information (as described in Lemma~\ref{Junelemma3}). Now, in the M-network part, let $e_{11}$ carry $a$, $e_{13}$ carry $b$, $e_{22}$ carry $x$, and $e_{23}$ carry $y$. Then, it can be easily seen that terminals $t_1,t_2$ and $t_3$ can retrieve its desired information. The terminal $t_4$ receives $a$ from $e_{11}$, $b$ from $(v_3,t_4)$, $a+y$ from $e_1$, and as a result it can deduce $y$ (by subtracting $a$ from $a+y$).
\end{IEEEproof}
\section{Proof of Lemmas~\ref{4Mlemma2}, \ref{4Mlemma3}, and \ref{4Mlemma4}}\label{app2}
We first develop some general equations that hold for the network $\mathcal{N}_2$. Let $f$ be the function that maps the sources and edges of the network $\mathcal{N}_2$ to the ground set $G$ of a discrete polymatrid $\mathbb{D}_2$ with rank function $\rho$ such that $\mathcal{N}_2$ has a $d$-dimensional VLNC solution over $\mathbb{F}_q$ if and only if $\mathbb{D}_2$ is representable over $\mathbb{F}_q$ and $f$ follows the conditions given in Theorem~\ref{DP}. Now let $\mathtt{g} = \rho \circ f$. Let the edge $(\bar{u}_i,\bar{v}_j)$ for $1\leq i \leq 3$ and $1\leq j\leq 5$ be denoted by $\bar{e}_{ij}$. 

From [D4] of Theorem~\ref{DP}, we have $\mathtt{g}(\bar{a},\bar{b},\bar{c}) = \mathtt{g}(\bar{a},\bar{b},\bar{c},\bar{e}_{11})$. From Lemma~\ref{toughlemma}, we also have $\mathtt{g}(\bar{a},\bar{b},\bar{c}) + \mathtt{g}(\bar{r},\bar{s},\bar{w}) + \mathtt{g}(\bar{x},\bar{y},\bar{z}) = \mathtt{g}(\bar{a},\bar{b},\bar{c},\bar{r},\bar{s},\bar{w},\bar{x},\bar{y},\bar{z})$. So this means $\mathtt{g}(\bar{a},\bar{b},\bar{c},\bar{e}_{11}) + \mathtt{g}(\bar{r},\bar{s},\bar{w},\bar{e}_{22}) + \mathtt{g}(\bar{x},\bar{y},\bar{z},\bar{e}_{33}) = \mathtt{g}(\bar{a},\bar{b},\bar{c},\bar{r},\bar{s},\bar{w},\bar{x},\bar{y},\bar{z},\\\bar{e}_{11},\bar{e}_{22},\bar{e}_{33})$. Now, we have:
\begin{IEEEeqnarray*}{l}
\mathtt{g}(\bar{e}_{11},\bar{a}) + \mathtt{g}(\bar{e}_{22},\bar{r}) + \mathtt{g}(\bar{e}_{33},\bar{x})\\
= \mathtt{g}(\bar{e}_{11},\bar{a},\bar{e}_{22},\bar{r},\bar{e}_{33},\bar{x}) \qquad\text{ [from Lemma~\ref{lema}]}\\
\leq \mathtt{g}(\bar{e}_{11},\bar{a},\bar{e}_{22},\bar{r},\bar{e}_{33},\bar{x}, (\bar{v}_4,\bar{t}_1), (\bar{v}_5,\bar{t}_1))\\
= \mathtt{g}(\bar{e}_{11},\bar{e}_{22},\bar{e}_{33},(\bar{v}_4,\bar{t}_1), (\bar{v}_5,\bar{t}_1)) \qquad\text{ [due to demands of $\bar{t}_1$]}\\
\leq 5d \qquad\text{ [from [D3] of Theorem~\ref{DP}]} \IEEEyesnumber\label{24f1}
\end{IEEEeqnarray*}
Similar to the equation (\ref{24f1}), we have the following equations:
\begin{IEEEeqnarray}{l}
\mathtt{g}(\bar{e}_{11},\bar{a}) + \mathtt{g}(\bar{e}_{22},\bar{r}) + \mathtt{g}(\bar{e}_{33},\bar{y}) \leq 5d\qquad \text{ [due to $\bar{t}_2$]}\label{24f2} \\
\mathtt{g}(\bar{e}_{11},\bar{a}) + \mathtt{g}(\bar{e}_{22},\bar{r}) + \mathtt{g}(\bar{e}_{33},\bar{z}) \leq 5d\qquad \text{ [due to $\bar{t}_3$]}\label{24f3} \\
\mathtt{g}(\bar{e}_{11},\bar{a}) + \mathtt{g}(\bar{e}_{22},\bar{s}) + \mathtt{g}(\bar{e}_{33},\bar{x}) \leq 5d\qquad \text{ [due to $\bar{t}_4$]}  \label{24f4}\\
\mathtt{g}(\bar{e}_{11},\bar{a}) + \mathtt{g}(\bar{e}_{22},\bar{s}) + \mathtt{g}(\bar{e}_{33},\bar{y}) \leq 5d\qquad \text{ [due to $\bar{t}_5$]}  \label{24f5} \\
\mathtt{g}(\bar{e}_{11},\bar{a}) + \mathtt{g}(\bar{e}_{22},\bar{s}) + \mathtt{g}(\bar{e}_{33},\bar{z}) \leq 5d\qquad \text{ [due to $\bar{t}_6$]}  \label{24f6} \\
\mathtt{g}(\bar{e}_{11},\bar{b}) + \mathtt{g}(\bar{e}_{22},\bar{r}) + \mathtt{g}(\bar{e}_{33},\bar{x}) \leq 5d\qquad \text{ [due to $\bar{t}_{10}$]}  \label{24f7} \\
\mathtt{g}(\bar{e}_{11},\bar{b}) + \mathtt{g}(\bar{e}_{22},\bar{r}) + \mathtt{g}(\bar{e}_{33},\bar{y}) \leq 5d\qquad \text{ [due to $\bar{t}_{11}$]}  \label{24f8} \\
\mathtt{g}(\bar{e}_{11},\bar{b}) + \mathtt{g}(\bar{e}_{22},\bar{r}) + \mathtt{g}(\bar{e}_{33},\bar{z}) \leq 5d\qquad \text{ [due to $\bar{t}_{12}$]}  \label{24f9} \\
\mathtt{g}(\bar{e}_{11},\bar{b}) + \mathtt{g}(\bar{e}_{22},\bar{s}) + \mathtt{g}(\bar{e}_{33},\bar{x}) \leq 5d\qquad \text{ [due to $\bar{t}_{13}$]} \label{24f10} \\
\mathtt{g}(\bar{e}_{11},\bar{b}) + \mathtt{g}(\bar{e}_{22},\bar{s}) + \mathtt{g}(\bar{e}_{33},\bar{y}) \leq 5d\qquad \text{ [due to $\bar{t}_{14}$]}  \label{24f11} \\
\mathtt{g}(\bar{e}_{11},\bar{b}) + \mathtt{g}(\bar{e}_{22},\bar{s}) + \mathtt{g}(\bar{e}_{33},\bar{z}) \leq 5d\qquad \text{ [due to $\bar{t}_{15}$]}  \label{24f12} \\
\mathtt{g}(\bar{e}_{11},\bar{c}) + \mathtt{g}(\bar{e}_{22},\bar{w}) + \mathtt{g}(\bar{e}_{33},\bar{z}) \leq 5d\qquad \text{ [due to $\bar{t}_{27}$]} \label{24f13}
\end{IEEEeqnarray}
It can be seen that due to terminals $\bar{t}_1, \bar{t}_{14}$, and $\bar{t}_{27}$, all of the source messages are to be retrieved from the edges in the set $\{\bar{e}_{ii},\bar{e}_{ij}|i=1,2,3; j=4,5 \}$. So we must have:
\begin{IEEEeqnarray*}{ll}
9d\, &= \mathtt{g}(\bar{a},\bar{b},\bar{c},\bar{r},\bar{s},\bar{w},\bar{x},\bar{y},\bar{z}) \qquad \text{ [from Lemma~\ref{toughlemma} and [D3] of Thm.~\ref{DP}]}\\
&= \mathtt{g}(\bar{a},\bar{b},\bar{c},\bar{r},\bar{s},\bar{w},\bar{x},\bar{y},\bar{z},\bar{e}_{11},\bar{e}_{14},\bar{e}_{15},\bar{e}_{22},\bar{e}_{24},\bar{e}_{25},\bar{e}_{33},\bar{e}_{34},\bar{e}_{35})\\
&= \mathtt{g}(\bar{e}_{11},\bar{e}_{14},\bar{e}_{15},\bar{e}_{22},\bar{e}_{24},\bar{e}_{25},\bar{e}_{33},\bar{e}_{34},\bar{e}_{35}) \qquad \text{ [due to demands of $\bar{t}_1, \bar{t}_{14}$, and $\bar{t}_{27}$]}\\
&\leq \sum_{i=1,2,3; j=i,4,5} \!\!\! \!\!\!\!\!\!\mathtt{g}(\bar{e}_{ij}) \qquad \text{ [from [P3] of Definition~\ref{P}]}\\
&\leq 9d \qquad \text{ [from [D3] of Theorem~\ref{DP}]}
\end{IEEEeqnarray*}
So we must have:
\begin{equation}
\mathtt{g}(\bar{e}_{ij}) = d \text{ for } i = 1,2,3 \text{ and } j=i,4,5.\label{hhh1}
\end{equation}
Now,
\begin{IEEEeqnarray*}{l}
\mathtt{g}(\bar{e}_{11},\bar{a}) + \mathtt{g}(\bar{e}_{11},\bar{b}) + \mathtt{g}(\bar{e}_{11},\bar{c})\\
\geq \mathtt{g}(\bar{e}_{11},\bar{a},\bar{b}) + \mathtt{g}(\bar{e}_{11}) + \mathtt{g}(\bar{e}_{11},\bar{c}) \qquad \text{ [applying [P3] of Definition~\ref{P}]}\\
\geq \mathtt{g}(\bar{e}_{11},\bar{a},\bar{b},\bar{c}) + 2\mathtt{g}(\bar{e}_{11})\qquad \text{ [applying [P3] of Definition~\ref{P}]}\\
= \mathtt{g}(\bar{a},\bar{b},\bar{c}) + 2\mathtt{g}(\bar{e}_{11})\\
= 5d \qquad \text{ [using equation~(\ref{hhh1})]}\IEEEyesnumber\label{24f14}\\
\end{IEEEeqnarray*}
Similarly, we also have the following inequalities:
\begin{IEEEeqnarray}{l}
\mathtt{g}(\bar{e}_{22},\bar{r}) + \mathtt{g}(\bar{e}_{22},\bar{s}) + \mathtt{g}(\bar{e}_{22},\bar{w}) \geq 5d\label{24f15}\\
\mathtt{g}(\bar{e}_{33},\bar{x}) + \mathtt{g}(\bar{e}_{33},\bar{y}) + \mathtt{g}(\bar{e}_{33},\bar{z}) \geq 5d\label{24f16}
\end{IEEEeqnarray}
%
%
Adding equations (\ref{24f14})-(\ref{24f16}), we have:
\begin{IEEEeqnarray*}{l}
\mathtt{g}(\bar{e}_{11},\bar{a}) + \mathtt{g}(\bar{e}_{11},\bar{b}) + \mathtt{g}(\bar{e}_{11},\bar{c}) + \mathtt{g}(\bar{e}_{22},\bar{r}) + \mathtt{g}(\bar{e}_{22},\bar{s}) + \mathtt{g}(\bar{e}_{22},\bar{w}) + \mathtt{g}(\bar{e}_{33},\bar{x}) \\ \hfill +\> \mathtt{g}(\bar{e}_{33},\bar{y}) + \mathtt{g}(\bar{e}_{33},\bar{z}) \geq 15d\\
\text{or, } (\mathtt{g}(\bar{e}_{11},\bar{a}) + \mathtt{g}(\bar{e}_{22},\bar{r}) + \mathtt{g}(\bar{e}_{33},\bar{x})) + (\mathtt{g}(\bar{e}_{11},\bar{b}) + \mathtt{g}(\bar{e}_{22},\bar{s}) + \mathtt{g}(\bar{e}_{33},\bar{y})) + \mathtt{g}(\bar{e}_{11},\bar{c}) \\ \hfill +\> \mathtt{g}(\bar{e}_{22},\bar{w}) + \mathtt{g}(\bar{e}_{33},\bar{z}) \geq 15d\IEEEyesnumber\IEEEeqnarraynumspace\label{24f17}
\end{IEEEeqnarray*}
From equations (\ref{24f1}), (\ref{24f11}), (\ref{24f13}), and (\ref{24f17}), we have:
\begin{IEEEeqnarray}{l}
\mathtt{g}(\bar{e}_{11},\bar{a}) + \mathtt{g}(\bar{e}_{22},\bar{r}) + \mathtt{g}(\bar{e}_{33},\bar{x}) = 5d\label{242f1}\\
\mathtt{g}(\bar{e}_{11},\bar{b}) + \mathtt{g}(\bar{e}_{22},\bar{s}) + \mathtt{g}(\bar{e}_{33},\bar{y}) = 5d\label{242f2}\\
\mathtt{g}(\bar{e}_{11},\bar{c}) + \mathtt{g}(\bar{e}_{22},\bar{w}) + \mathtt{g}(\bar{e}_{33},\bar{z}) = 5d\label{242f3}
\end{IEEEeqnarray}
Rearranging equation (\ref{24f17}) and then using equations (\ref{24f2}), (\ref{24f10}), and (\ref{24f13}), we have:
\begin{IEEEeqnarray}{l}
\mathtt{g}(\bar{e}_{11},\bar{a}) + \mathtt{g}(\bar{e}_{22},\bar{r}) + \mathtt{g}(\bar{e}_{33},\bar{y}) = 5d\label{242f4}\\
\mathtt{g}(\bar{e}_{11},\bar{b}) + \mathtt{g}(\bar{e}_{22},\bar{s}) + \mathtt{g}(\bar{e}_{33},\bar{x}) = 5d\label{242f9}
\end{IEEEeqnarray}
Rearranging equation (\ref{24f17}) and then using equations (\ref{24f4}), (\ref{24f8}), and (\ref{24f13}), we have:
\begin{IEEEeqnarray}{l}
\mathtt{g}(\bar{e}_{11},\bar{a}) + \mathtt{g}(\bar{e}_{22},\bar{s}) + \mathtt{g}(\bar{e}_{33},\bar{x}) = 5d\label{242f5}\\
\mathtt{g}(\bar{e}_{11},\bar{b}) + \mathtt{g}(\bar{e}_{22},\bar{r}) + \mathtt{g}(\bar{e}_{33},\bar{y}) = 5d\label{242f8}
\end{IEEEeqnarray}
Rearranging equation (\ref{24f17}) and then using equations (\ref{24f5}), (\ref{24f7}), and (\ref{24f13}), we have:
\begin{IEEEeqnarray}{l}
\mathtt{g}(\bar{e}_{11},\bar{a}) + \mathtt{g}(\bar{e}_{22},\bar{s}) + \mathtt{g}(\bar{e}_{33},\bar{y}) = 5d\label{242f6}\\
\mathtt{g}(\bar{e}_{11},\bar{b}) + \mathtt{g}(\bar{e}_{22},\bar{r}) + \mathtt{g}(\bar{e}_{33},\bar{x}) = 5d\label{242f7}
\end{IEEEeqnarray}
Subtracting equations (\ref{242f1}) from (\ref{242f7}), we get:
\begin{equation}
\mathtt{g}(\bar{e}_{11},\bar{a}) = \mathtt{g}(\bar{e}_{11},\bar{b})\label{243f6}
\end{equation}
Subtracting equations (\ref{242f1}) from (\ref{242f5}), we get:
\begin{equation}
\mathtt{g}(\bar{e}_{22},\bar{r}) = \mathtt{g}(\bar{e}_{22},\bar{s})\label{243f7}
\end{equation}
Subtracting equations (\ref{242f1}) from (\ref{242f4}), we get:
\begin{equation}
\mathtt{g}(\bar{e}_{33},\bar{x}) = \mathtt{g}(\bar{e}_{33},\bar{y})\label{243f8}
\end{equation}
Adding equations (\ref{242f1}), (\ref{242f2}) and (\ref{242f3}), we have:
\begin{IEEEeqnarray}{l}
(\mathtt{g}(\bar{e}_{11},\bar{a}) + \mathtt{g}(\bar{e}_{11},\bar{b}) + \mathtt{g}(\bar{e}_{11},\bar{c})) + (\mathtt{g}(\bar{e}_{22},\bar{r}) + \mathtt{g}(\bar{e}_{22},\bar{s}) + \mathtt{g}(\bar{e}_{22},\bar{w})) + (\mathtt{g}(\bar{e}_{33},\bar{x})\IEEEnonumber \\\hfill +\> \mathtt{g}(\bar{e}_{33},\bar{y}) + \mathtt{g}(\bar{e}_{33},\bar{z})) = 15d\IEEEeqnarraynumspace\label{243f2}
\end{IEEEeqnarray}
As equations (\ref{24f14})-(\ref{24f16}) hold, we must have:
\begin{IEEEeqnarray}{l}
\mathtt{g}(\bar{e}_{11},\bar{a}) + \mathtt{g}(\bar{e}_{11},\bar{b}) + \mathtt{g}(\bar{e}_{11},\bar{c}) = 5d\label{243f3}\\
\mathtt{g}(\bar{e}_{22},\bar{r}) + \mathtt{g}(\bar{e}_{22},\bar{s}) + \mathtt{g}(\bar{e}_{22},\bar{w}) = 5d\label{243f4}\\
\mathtt{g}(\bar{e}_{33},\bar{x}) + \mathtt{g}(\bar{e}_{33},\bar{y}) + \mathtt{g}(\bar{e}_{33},\bar{z}) = 5d\label{243f5}
\end{IEEEeqnarray}
Applying equations (\ref{243f6})-(\ref{243f8}) to equations (\ref{243f3})-(\ref{243f5}), we have:
\begin{IEEEeqnarray}{l}
2\mathtt{g}(\bar{e}_{11},\bar{a}) + \mathtt{g}(\bar{e}_{11},\bar{c}) = 5d\label{244f1}\\
2\mathtt{g}(\bar{e}_{22},\bar{r}) + \mathtt{g}(\bar{e}_{22},\bar{w}) = 5d\label{244f2}\\
2\mathtt{g}(\bar{e}_{33},\bar{x}) + \mathtt{g}(\bar{e}_{33},\bar{z}) = 5d\label{244f3}
\end{IEEEeqnarray}
Multiplying equation (\ref{24f3}) by $2$ and then adding to equation (\ref{242f3}), we have:
\begin{IEEEeqnarray*}{l}
2(\mathtt{g}(\bar{e}_{11},\bar{a}) + \mathtt{g}(\bar{e}_{22},\bar{r}) + \mathtt{g}(\bar{e}_{33},\bar{z})) + \mathtt{g}(\bar{e}_{11},\bar{c}) + \mathtt{g}(\bar{e}_{22},\bar{w}) + \mathtt{g}(\bar{e}_{33},\bar{z}) \leq 15d\\
\text{or, } 5d + 5d + 3\mathtt{g}(\bar{e}_{33},\bar{z}) \leq 15d \qquad\text{ [substituting equations (\ref{244f1}) and (\ref{244f2})]}\\
\text{or, } 3\mathtt{g}(\bar{e}_{33},\bar{z}) \leq 5d\\
\text{or, } \mathtt{g}(\bar{e}_{33},\bar{z}) \leq \frac{5d}{3}\IEEEyesnumber\label{245f1}
\end{IEEEeqnarray*}
We now derive one more equation that must hold if the characteristic of the finite field does not divide $q^\prime $. Note that in such a case, from Lemma~\ref{july63}, we have: $\mathtt{g}(\bar{a}) = \mathtt{g}(\bar{a},\bar{e}_1)$. So due to the demands of terminal $\bar{t}_{25}$, we have:
\begin{IEEEeqnarray*}{l}
\mathtt{g}(\bar{e}_{11},\bar{a},\bar{c}) + \mathtt{g}(\bar{e}_{22},\bar{w}) + \mathtt{g}(\bar{e}_{33},\bar{x}) \\
= \mathtt{g}(\bar{e}_{11},\bar{e}_{22},\bar{e}_{33},\bar{a},\bar{c},\bar{w},\bar{x}) \qquad\text{ [from Lemma~\ref{lema}]}\\
\leq \mathtt{g}(\bar{e}_{11},\bar{e}_{22},\bar{e}_{33},\bar{a},\bar{c},\bar{w},\bar{x},(\bar{v}_4,\bar{t}_{25}), (\bar{v}_5,\bar{t}_{25}))\\
= \mathtt{g}(\bar{e}_{11},\bar{e}_{22},\bar{e}_{33},\bar{a},\bar{c},\bar{w},\bar{x},(\bar{v}_4,\bar{t}_{25}), (\bar{v}_5,\bar{t}_{25}),\bar{e}_1) \qquad\text{ [as $\mathtt{g}(\bar{a}) = \mathtt{g}(\bar{a},\bar{e}_1)$ from Lemma~\ref{july63}]}\\
= \mathtt{g}(\bar{e}_{11},\bar{e}_{22},\bar{e}_{33},\bar{a},(\bar{v}_4,\bar{t}_{25}), (\bar{v}_5,\bar{t}_{25}),\bar{e}_1) \qquad\text{ [due to demands of $\bar{t}_{25}$]}\\
= \mathtt{g}(\bar{e}_{11},\bar{e}_{22},\bar{e}_{33},\bar{a},(\bar{v}_4,\bar{t}_{25}), (\bar{v}_5,\bar{t}_{25}))\\
\leq \mathtt{g}(\bar{e}_{22},\bar{e}_{33},(\bar{v}_4,\bar{t}_{25}), (\bar{v}_5,\bar{t}_{25})) + \mathtt{g}(\bar{e}_{11},\bar{a})\\
\leq 4d + \mathtt{g}(\bar{e}_{11},\bar{a})\IEEEyesnumber\label{244f4}
\end{IEEEeqnarray*}
%
%
%
Also,
\begin{IEEEeqnarray*}{l}
\mathtt{g}(\bar{e}_{11},\bar{a},\bar{c}) + \mathtt{g}(\bar{e}_{11},\bar{a}) \\
= \mathtt{g}(\bar{e}_{11},\bar{a},\bar{c}) + \mathtt{g}(\bar{e}_{11},\bar{b}) \qquad\text{[from equation (\ref{243f6})]}\\
\geq \mathtt{g}(\bar{e}_{11},\bar{a},\bar{c},\bar{b}) + \mathtt{g}(\bar{e}_{11}) \qquad \text{[using [P3] of Definition~\ref{P}]}\\
= 4d \qquad\text{ [using equation~(\ref{hhh1})]}
\end{IEEEeqnarray*}
Then, we have:
\begin{equation}
\mathtt{g}(\bar{e}_{11},\bar{a},\bar{c}) \geq 4d - \mathtt{g}(\bar{e}_{11},\bar{a})\IEEEyesnumber\label{244f5}
\end{equation}
Substituting equation (\ref{244f5}) in equation (\ref{244f4}), we have:
\begin{equation}
\mathtt{g}(\bar{e}_{11},\bar{w}) + \mathtt{g}(\bar{e}_{33},\bar{x}) \leq 2\mathtt{g}(\bar{e}_{11},\bar{a})\label{244f6}
\end{equation}
\subsection{Proof of Lemma~\ref{4Mlemma2}}\label{app2.1}
\begin{IEEEproof}
Note that equations (\ref{244f4})-(\ref{244f6}) cannot be used as they hold if the characteristic of the finite field does not divide $q^\prime $; and this lemma has to be shown to be true over all finite fields. Let us assume that the network has an SLNC solution. 

Since $d=1$, and the rank function of a discrete polymatroid is always an integer, from equation (\ref{245f1}), we have: $\mathtt{g}(\bar{e}_{33},\bar{z}) \leq 1$. Then from [P2] of Definition~\ref{P} and [D3] of Theorem~\ref{DP}, we have: 
\begin{equation}
\mathtt{g}(\bar{e}_{33},\bar{z}) = 1\label{245f2}
\end{equation}
Substituting equation (\ref{245f2}) in equation (\ref{242f3}), we have:
\begin{equation}
\mathtt{g}(\bar{e}_{11},\bar{c}) + \mathtt{g}(\bar{e}_{22},\bar{w}) = 4\label{245f3}
\end{equation}
Since rank of any element is less than or equal to $1$, we have: $\mathtt{g}(\bar{e}_{11},\bar{c}) \leq 2$ and $\mathtt{g}(\bar{e}_{22},\bar{w}) \leq 2$. Then equation (\ref{245f3}) implies:
\begin{equation}
\mathtt{g}(\bar{e}_{11},\bar{c}) = 2\label{245f4}
\end{equation}
Substituting equation (\ref{245f4}) in equation (\ref{244f1}), we have:
\begin{IEEEeqnarray}{l}
2\mathtt{g}(\bar{e}_{11},\bar{a}) = 3\IEEEnonumber\\
\text{or, } \mathtt{g}(\bar{e}_{11},\bar{a}) = 3/2\label{245f5}
\end{IEEEeqnarray}
Equation (\ref{245f5}) is a contradiction as by Definition~\ref{P} the rank function always outputs an integer.
\end{IEEEproof}
\subsection{Proof of Lemma~\ref{4Mlemma3}}\label{app2.2}
\begin{IEEEproof}
\begin{figure}
\centering
\includegraphics[width=\textwidth]{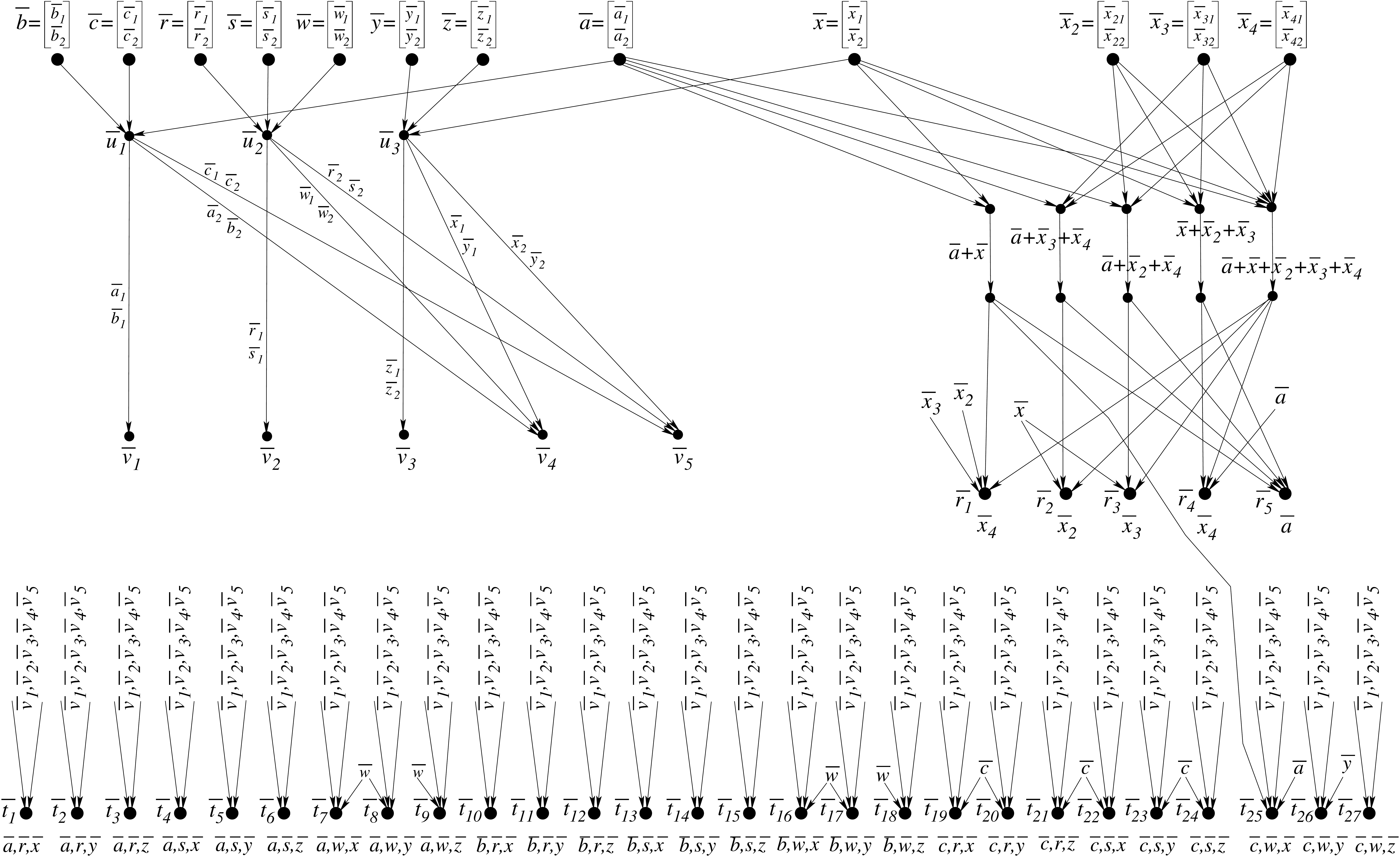}
\caption{A $2$-dimensional VLNC solution of the network $\mathcal{N}_2$ when $q^\prime = 2$ and the characteristic of the finite field divides $q^\prime$.}
\label{G2dim2}
\end{figure}
Consider the `only if' part. We show that if the characteristic of the finite field does not divide $q^\prime$ then network $\mathcal{N}_2$ has no $2$-dimensional VLNC solution. We prove this result by contradiction. Assume that $\mathcal{N}_2$ has a $2$-dimensional VLNC solution even when the characteristic of the finite field does not divide $q^\prime$.

Since the rank function of a discrete polymatroid is integer valued, from equation (\ref{245f1}), we have:
\begin{equation}
\mathtt{g}(\bar{e}_{33},\bar{z}) \leq 3\label{246f1}
\end{equation}
Substituting equation (\ref{246f1}) in equation (\ref{244f3}), we have:
\begin{equation}
\mathtt{g}(\bar{e}_{33},\bar{x}) \geq 3.5
\end{equation}
Then it must be that
\begin{equation}
\mathtt{g}(\bar{e}_{33},\bar{x}) \geq 4
\end{equation}
Since rank of an element is less than or equal to $2$ ([D3] of Theorem~\ref{DP}), we must have:
\begin{equation}
\mathtt{g}(\bar{e}_{33},\bar{x}) = 4\label{246f2}
\end{equation}
Substituting equation (\ref{246f2}) in equation (\ref{244f3}), we have:
\begin{equation}
\mathtt{g}(\bar{e}_{33},\bar{z}) = 2\label{246f3}
\end{equation}
Substituting equation (\ref{246f3}) in equation (\ref{242f3}), we have:
\begin{equation}
\mathtt{g}(\bar{e}_{11},\bar{c}) + \mathtt{g}(\bar{e}_{22},\bar{w}) = 8
\end{equation}
Since rank of an element is less than or equal to $2$, we must have:
\begin{IEEEeqnarray}{l}
\mathtt{g}(\bar{e}_{11},\bar{c}) = 4\label{246f4}\\
\mathtt{g}(\bar{e}_{22},\bar{w}) = 4\label{246f5}
\end{IEEEeqnarray}
Substituting equation (\ref{246f4}) in equation (\ref{244f1}), we have:
\begin{equation}
\mathtt{g}(\bar{e}_{11},\bar{a}) = 3\label{246f6}
\end{equation}
Substituting equations (\ref{246f2}), (\ref{246f5}), and (\ref{246f6}) in equation (\ref{244f6}), we have: $8 \leq 6$, which is a contradiction.

To prove the `if' part we present a $2$-dimensional VLNC solution over a finite field whose characteristic divides $q^\prime$. Let the message vector generated by a source be denoted by the same label as the source. In Fig.~\ref{G2dim2} we show a $2$-dimensional VLNC solution of $\mathcal{N}_2$ when $q^\prime = 2$ (see Lemma~\ref{Junelemma3} for the decoding operations at the terminals of the Char-$q^\prime$-$\bar{x}$ sub-network). This solution can easily be extended for any value of $q^\prime$. (For a different value of $q^\prime$, only the decoding matrices at the terminals of the Char-$q^\prime$-$\bar{x}$ sub-network changes (see equation (\ref{4ap1}).) 

\end{IEEEproof}
\subsection{Proof of Lemma~\ref{4Mlemma4}}\label{app2.3}
\begin{IEEEproof}
Consider the `only if' part. We show that if the characteristic of the finite field does not divide $q^\prime $ then the network $\mathcal{N}_2$ has no $5$-dimensional VLNC solution. We prove this result by contradiction. Assume that $\mathcal{N}_2$ has a $5$-dimensional VLNC solution even when the characteristic of the finite field does not divide $q^\prime $.

Since the rank function of a discrete polymatroid is integer valued, from equation (\ref{245f1}), we have:
\begin{equation}
\mathtt{g}(\bar{e}_{33},\bar{z}) \leq 8\label{247f1}
\end{equation}
From  equation (\ref{244f3}), we get that $25 - \mathtt{g}(\bar{e}_{33},\bar{z})$ must be divisible by $2$ (otherwise $\mathtt{g}(\bar{e}_{33},\bar{x})$ would not be an integer). Hence $\mathtt{g}(\bar{e}_{33},\bar{z})$ must be an odd number. Using similar reasoning, from equations (\ref{244f1}) and (\ref{244f2}), we get that $\mathtt{g}(\bar{e}_{11},\bar{c})$ and $\mathtt{g}(\bar{e}_{22},\bar{w})$ must be odd numbers.

Then, since $5 = \mathtt{g}(\bar{z}) \leq \mathtt{g}(\bar{e}_{33},\bar{z})$, either $\mathtt{g}(\bar{e}_{33},\bar{z}) = 5$ or $\mathtt{g}(\bar{e}_{33},\bar{z}) = 7$.

\noindent \textbf{Case I}: $\mathtt{g}(\bar{e}_{33},\bar{z}) = 5$.

Substituting $\mathtt{g}(\bar{e}_{33},\bar{z}) = 5$ in equation (\ref{242f3}), we get:
\begin{equation}
\mathtt{g}(\bar{e}_{11},\bar{c}) + \mathtt{g}(\bar{e}_{22},\bar{w}) = 20
\end{equation}
Since rank of any union of two elements is less than or equal to $10$, we must have
\begin{equation}
\mathtt{g}(\bar{e}_{11},\bar{c}) = \mathtt{g}(\bar{e}_{22},\bar{w}) = 10\label{247f2}
\end{equation}
But equation (\ref{247f2}) is a contradiction because as we have argued, $\mathtt{g}(\bar{e}_{11},\bar{c})$ and $\mathtt{g}(\bar{e}_{22},\bar{w})$ must be odd numbers.

\noindent \textbf{Case II}: $\mathtt{g}(\bar{e}_{33},\bar{z}) = 7$.

Substituting $\mathtt{g}(\bar{e}_{33},\bar{z}) = 7$ in equation (\ref{244f3}), we have:
\begin{equation}
\mathtt{g}(\bar{e}_{33},\bar{x}) = 9\label{247f3}
\end{equation}
Substituting $\mathtt{g}(\bar{e}_{33},\bar{z}) = 7$ in equation (\ref{242f3}), we get:
\begin{equation}
\mathtt{g}(\bar{e}_{11},\bar{c}) + \mathtt{g}(\bar{e}_{22},\bar{w}) = 18
\end{equation}
Since neither of $\mathtt{g}(\bar{e}_{11},\bar{c})$ and $\mathtt{g}(\bar{e}_{22},\bar{w})$ can be equal to $10$ (as $10$ is an even number), and as both of them must be less than $10$, we must have:
\begin{IEEEeqnarray}{l}
\mathtt{g}(\bar{e}_{11},\bar{c}) = 9\label{247f4}\\
\mathtt{g}(\bar{e}_{22},\bar{w}) = 9\label{247f5}
\end{IEEEeqnarray}
Substituting equation (\ref{247f4}) in equation (\ref{244f1}), we have:
\begin{equation}
\mathtt{g}(\bar{e}_{11},\bar{a}) = 8\label{247f6}\\
\end{equation}
Substituting equations (\ref{247f3}), (\ref{247f5}), and (\ref{247f6}) in equation (\ref{244f6}), we have: $18 \leq 16$, which is a contradiction.

To prove the `if' part we now show a $5$-dimensional VLNC solution when the characteristic of the finite field divides $q^\prime $. From Lemma~\ref{Julylemma6}, we know that $\mathcal{N}_2$ has a $3$-dimensional VLNC solution over all finite fields. From Lemma~\ref{4Mlemma3}, we know that $\mathcal{N}_2$ has a $2$-dimensional VLNC solution over a finite field whose characteristic divides $q^\prime$. So a $5$-dimensional VLNC solution over a finite field whose characteristic divides $q^\prime$ can easily be constructed.
\end{IEEEproof}

%
%
%
%
%
%
%
%

\ifCLASSOPTIONcaptionsoff
  \newpage
\fi

\end{document}